\documentclass[12pt,preprint]{aastex}

\usepackage{natbib}
\def\degree{\ifmmode {^\circ}\else {$^\circ$}\fi}
\def\rstar{\ifmmode {\, R_{\star}}\else $R_{\star}$\fi}
\def\msol{\ifmmode {\, M_{\odot}}\else $M_{\odot}$\fi}
\def\rsol{\ifmmode {\, R_{\odot}}\else $R_{\odot}$\fi}
\def\lsol{\ifmmode {\, L_{\odot}}\else $L_{\odot}$\fi}
\def\msolyr{\ifmmode {\, M_{\odot}\,{\rm yr}^{-1}}\else $M_{\odot}\,{\rm yr}^{-1}$\fi}
\def\mdot{\ifmmode {\,\dot{M}}\else $\dot{M}$\fi}
\def\mdotyr{\ifmmode {\,\dot{M}\,yr^{-1}}\else $\dot{M}\,yr^{-1}$\fi}

\newcommand{\Teff}{\ifmmode{T_{\rm eff}}\else{$T_{\rm eff}$}}

\begin{document}

\title{The Circumstellar Environment of  R Coronae Borealis: White Dwarf Merger or Final Helium Shell Flash? \footnote{Based in part on observations with the NASA/ESA {\it Hubble Space Telescope}
obtained at the Space Telescope Science Institute, and from the data archive at
STScI, which are operated by the Association of Universities for Research in
Astronomy, Inc., under NASA contract NAS5-26555}}


\author{Geoffrey C. Clayton\altaffilmark{1}, Ben E.K. Sugerman\altaffilmark{2}, S. Adam Stanford\altaffilmark{3}, B. A. Whitney\altaffilmark{4,5}, J. Honor\altaffilmark{5},  B. Babler\altaffilmark{5}, M.J. Barlow\altaffilmark{6}, K.D. Gordon\altaffilmark{7}, J.E. Andrews\altaffilmark{1}, T.R. Geballe\altaffilmark{8}, Howard E. Bond\altaffilmark{7}, O. De Marco\altaffilmark{9}, W.A. Lawson\altaffilmark{10}, B. Sibthorpe\altaffilmark{11},  G. Olofsson\altaffilmark{12}, E. Polehampton\altaffilmark{13,14}, H. L. Gomez\altaffilmark{15}, M. Matsuura\altaffilmark{7,16},  P. C. Hargrave\altaffilmark{15},  R. J. Ivison\altaffilmark{17}, R. Wesson\altaffilmark{7}, S. J. Leeks\altaffilmark{13}, B. M. Swinyard\altaffilmark{13}, and T. L. Lim\altaffilmark{13}}

\altaffiltext{1}{Dept.\ of Physics \& Astronomy, Louisiana State
University, Baton Rouge, LA 70803; gclayton, jandrews@phys.lsu.edu}
\altaffiltext{2}{Dept.\ of Physics and Astronomy, Goucher College, 1021 Dulaney Valley Rd., Baltimore, MD 21204; ben.sugerman@goucher.edu}
\altaffiltext{3}{IGPP, Lawrence Livermore National Laboratory, Livermore, CA 94551; stanford@physics.ucdavis.edu}
\altaffiltext{4}{Space Science Institute, 4750 Walnut St. Suite 205, Boulder, CO
80301; bwhitney@spacescience.org}
\altaffiltext{5}{Dept.\ of Astronomy, 475 North Charter St., University of
Wisconsin, Madison, WI 53706; jhonor, brian@astro.wisc.edu}
\altaffiltext{6}{Dept.\ of Physics and Astronomy, UCL, Gower Street, London WC1E 6BT, UK; mjb, mikako, rwesson@star.ucl.ac.uk}
\altaffiltext{7}{STScI, 3700 San Martin Dr., Baltimore, MD 21218; bond, gordon@stsci.edu}
\altaffiltext{8}{Gemini Observatory, 670 N. A'ohoku Place, Hilo, HI 96720;  
tgeballe@gemini.edu}
\altaffiltext{9}{Dept.\ of Physics, Macquarie University, Sydney, NSW 2109, Australia; orsola@science.mq.edu.au}
\altaffiltext{10}{School of PEMS, University of New South Wales, ADFA, P.O. Box 7916, Canberra 2610, Australia; w.lawson@adfa.edu.au}
\altaffiltext{11}{Max-Planck-Institut f\"{u}r Astronomie, K\"{o}nigstuhl 17, 69117 Heidelberg, Germany;\\ bruce.sibthorpe@stfc.ac.uk}
\altaffiltext{12}{Dept.\ of Astronomy, Stockholm University, AlbaNova University Center, Roslagstullsbacken 21, 10691 Stockholm, Sweden; olofsson@astro.su.se}
\altaffiltext{13}{Space Science and Technology Dept., Rutherford
Appleton Laboratory, Oxfordshire, OX11 0QX, UK;
 edward.polehampton, sarah.leeks, bruce.swinyard,
  tanya.lim@stfc.ac.uk}
\altaffiltext{14}{Institute for Space Imaging Science, University of Lethbridge, Lethbridge, Alberta, T1J 1B1, Canada}
\altaffiltext{15}{School of Physics and Astronomy, Cardiff University, 5 The Parade, Cardiff, Wales CF24 3YB, UK; Haley.Gomez, p.hargrave}
\altaffiltext{16}{MSSL, UCL, Holmbury St.\ Mary, Dorking, Surrey RH5 6NT, UK}
\altaffiltext{17}{UK Astronomy Technology Centre, ROE, Blackford Hill, Edinburgh EH9 3HJ, UK; rji@roe.ac.uk}

\begin{abstract}
In 2007, R Coronae Borealis (R~CrB) went into an historically deep and
long decline. In this state, the dust acts like a natural coronagraph at visible
wavelengths, allowing faint
nebulosity around the star to be seen.  Imaging has been obtained from 0.5 to 500 \micron~with
{\it Gemini/GMOS}, {\it HST/WFPC2}, {\it Spitzer/MIPS}, and {\it
  Herschel/SPIRE}.
Several of the structures around R~CrB are cometary globules caused by
wind from the star streaming past dense blobs.  The estimated dust mass
of the knots is consistent with their
being responsible for the R~CrB declines if they form along the line
of sight to the star. In addition, there is a large diffuse shell
extending up to 4 pc away from the star containing cool 25 K dust that
is detected all the way out to 500 \micron. The SED of R~CrB can be well fit by a 150
AU disk surrounded by a very large diffuse envelope which corresponds
to the size of the observed nebulosity. The total masses of the disk
and envelope are 10$^{-4}$ and 2 M$_{\sun}$, respectively, assuming a
gas-to-dust ratio of 100.
The evidence pointing toward a white-dwarf merger or a
final-helium-shell flash origin for R~CrB is contradictory.  The
shell and the
cometary knots are consistent with a fossil planetary nebula. Along with the fact that R~CrB shows
significant Lithium in its atmosphere, this supports the final-helium-shell flash. However, the relatively high
inferred mass of R~CrB and its high fluorine abundance support a white-dwarf merger.
\end{abstract}

\section{Introduction}

 R Coronae Borealis (R~CrB) was one of the first variable stars identified and its brightness variations have been monitored since its discovery over 200 years ago
\citep{1797RSPT...87..133P}.
The RCB stars form a small group of carbon-rich supergiants which
are defined by extreme hydrogen deficiency, and unusual variability, characterized by large declines of 8 mag or more due to the formation of carbon dust at irregular intervals \citep{1996PASP..108..225C}.
Two evolutionary scenarios have been suggested, a merger of a double-degenerate white dwarf merger, or a
final helium shell flash \citep[e.g.,][]{1984ApJ...277..355W, Clayton:2007ve}.
The unexpected discovery that some of the RCB stars have isotopic abundances of $^{18}$O which are up to 500 times those seen in other stars has increased the likelihood that these rare stars are the result of a white-dwarf merger \citep{2005ApJ...623L.141C,Clayton:2007ve,2010ApJ...714..144G}.

RCB stars have long been known to show IR excesses due to warm circumstellar dust \citep[and references therein]{1996PASP..108..225C,Feast:1997lr}.
Spectropolarimetry of R~CrB obtained in a deep decline showed optical depth effects with wavelength that are consistent with a bipolar geometry with a thick disk or torus obscuring the star and additional more diffuse dust around the poles \citep{1988ApJ...325L...9S,1997ApJ...476..870C}.
Resolved nebulosity has been detected from the UV to the IR around several RCB stars, including R~CrB itself, ranging in size from a few arcseconds (V854 Cen, UW Cen) to arcminutes (R~CrB, RY Sgr, V CrA, UW Cen)  \citep{Walker:1985rr,1986MNRAS.222..357K,1986ASSL..128..407W,1991MNRAS.248P...1P,1999ApJ...517L.143C,2001ApJ...560..986C},
possibly linking these stars with final helium shell flash stars and the central stars of planetary nebulae (PNe).
In particular, R~CrB has a far-IR shell which is $\sim$20\arcmin~in diameter, corresponding to 8 pc at its assumed distance \citep{Gillett:1986cr}. 
More recent observations have used interferometry and adaptive optics to obtain information about the distribution of dust around RCB stars on small spatial scales of tens of AU  \citep{2003A&A...408..553O,de-Laverny:2004lr,2007A&A...466L...1L,2011MNRAS.414.1195B}


In 2007 July, R~CrB went into a historically deep and long decline
reaching V=15. In this state, R~CrB acts as a natural coronagraph, as
the thick dust cloud in front of the star eclipses its photosphere but not its circumstellar material.
In this paper, we present new {\it Gemini/GMOS}, {\it HST/WFPC2}, {\it Spitzer/MIPS}, and {\it Herschel/SPIRE} images of R~CrB obtained just before or during this decline, giving wavelength coverage from 0.5 to 500 \micron. We will investigate the morphology and nature of the dust clouds seen near R~CrB as well as its large far-IR shell.

\section{Observations and Data Reduction}

\begin{figure}
\begin{center}
\includegraphics[width=5in,angle=0]{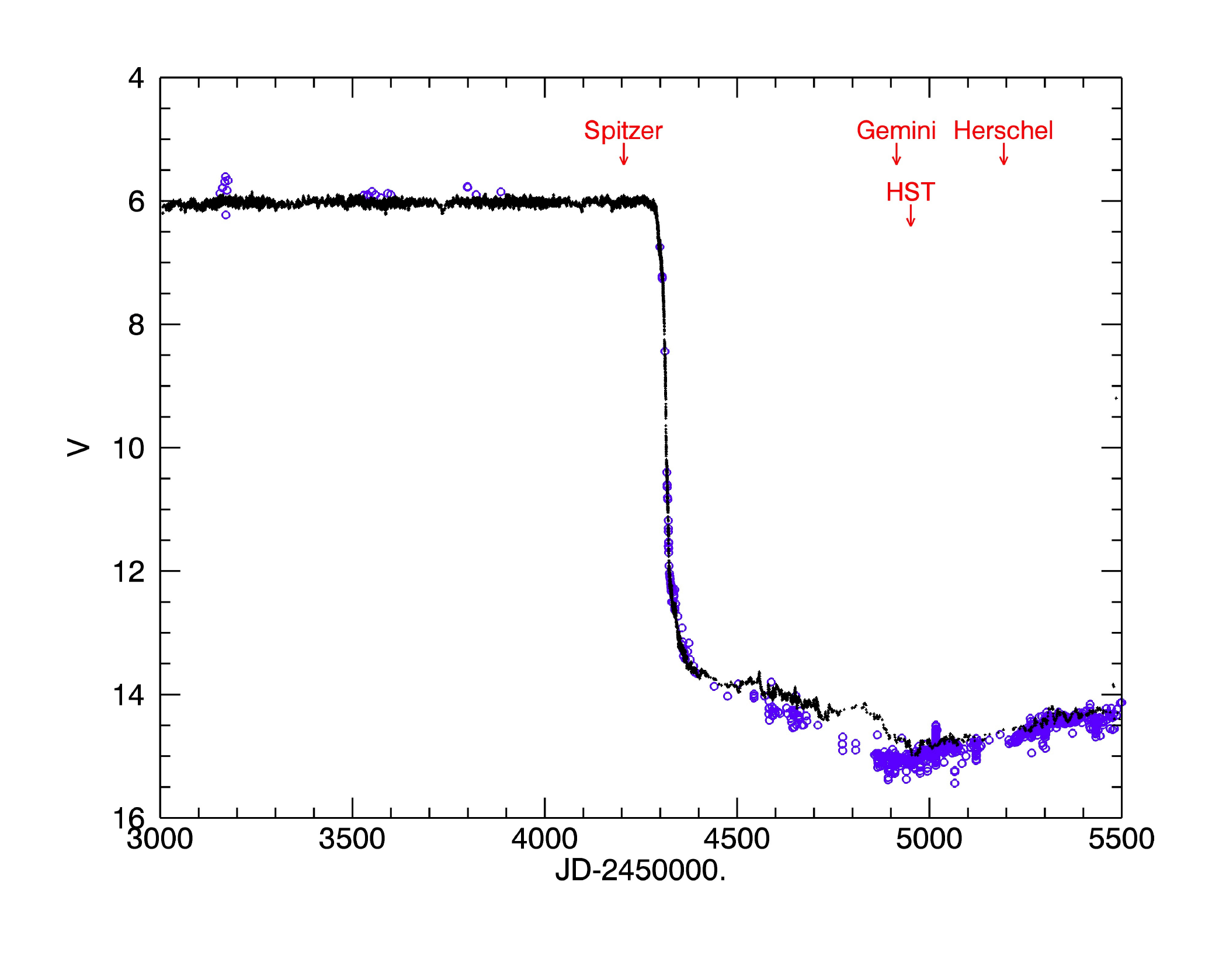}
\end{center}
\caption{AAVSO data for R~CrB since 2004. Visual magnitudes are plotted as black dots. Johnson V data are plotted as blue open circles. The epochs at which the {\it Spitzer}, {\it Gemini}, {\it HST} and {\it Herschel} data were obtained are marked. \label{fig1}}
\end{figure}

The new observations obtained for this study are summarized in Table 1. 
Figure \ref{fig1} shows the recent AAVSO lightcurve of R~CrB with the epochs of the various observations marked. 

R~CrB was observed with {\it Gemini-S/GMOS} on 2009 March 23. Images were obtained in the g\arcmin, r\arcmin, i\arcmin, z\arcmin, and CaT filters. 
We obtained 8 integrations of 30.5 s in each of the g\arcmin~and r\arcmin~filters for a total of 244s, and 14 integrations of 30.5 s each in the i\arcmin, z\arcmin, and CaT filters for a total of 427s. We used the 2x2 binning and slow readout mode. The field is 6\arcmin x 6\arcmin. The central part of the field is shown in Figure \ref{fig2}. 
All images have been geometrically registered to within 0.1 pix rms.  A point spread function (PSF) centered on R~CrB has been subtracted where the sky brightness has been ramped down to zero as one goes out 27 pixels in radius. 
The plate scale is 0\farcs146 per pixel. The seeing was $\sim$0\farcs8.
These PSF-subtracted images are shown in Figure \ref{fig3}. 

R~CrB was observed with {\it HST/WFPC2} on 2009 April 27. Images were obtained with the F555W and F814W filters with total integration times of 487s and 496s, respectively. The individual images were stacked and drizzled. 
The images were then registered and a PSF was subtracted. These images are shown in Figure \ref{fig5}.
WFPC2 images of R~CrB taken on 1996 July 1 (JD 2450265) with the F469N filter were downloaded from the MAST archive \citep{2000ApJ...528..861U}. The exposures were 4 x 20s and 2 x 400s. When these images were obtained R~CrB was at a visual magnitude  of $\sim$8.1 (AAVSO) while recovering toward maximum light after a deep decline. The individual images were stacked and drizzled.

 A UBVRI photometric sequence has been established for the R~CrB field \citep{2010PASP..122..541L}. A number of stars (Nos. 13, 14, 19, 20, 22, 23, 31, 34, 35, 37, 39) from that sequence are also in the {\it Gemini/GMOS} field. Two of the stars (Nos. 20, 22) lie in the {\it HST/WFPC2} field but none are on the PC chip. In addition, a strip from the Sloan Digital Sky Survey (SDSS) telescope crosses R~CrB. Several of the photometric sequence stars (Nos. 13, 14, 19, 20, 23, 35, 37, 39)  from \citet{2010PASP..122..541L} also have ugriz photometry\footnote{http://www.sdss.org/} and lie in the GMOS field. 

{\it Spitzer/MIPS} scan maps were obtained of R~CrB at 24, 70, and 160 \micron~on 2007 April 14/15 as part of program 30029. We mapped a 20\arcmin~wide area using 15 legs at the medium scan rate with 80\arcsec~(1/4 array) cross-scan offsets. The total integration time was 15,536s. 
R~CrB was observed
twice to allow for good Germanium (70 and 160 \micron) transient
and asteroid removal.  The second epoch maps allow for the
clean removal of the Germanium responsivity  drifts which are especially
important for mapping extended emission around R~CrB.
The scan maps were reduced and mosaiced using the MIPS Data
Analysis Tool \citep{2005PASP..117..503G} supplemented with custom reduction
scripts written specifically to improve the MIPS reductions of
extended sources.  The custom reduction scripts include extra stpdf beyond
that of the MIPS DAT.  At 24~$\mu$m, the extra stpdf include readout
offset correction, array averaged background subtraction (using a low
order polynomial fit to each leg, with the region including the Galaxy
excluded from this fit), and exclusion of the 1st five images in each
scan leg due to boost frame transients.  At 70 and 160~$\mu$m, the
extra processing step is a pixel dependent background subtraction for
each map to remove residual detector
drifts and background cirrus and zodiacal emissions.
The 24 \micron~image was saturated in the core and was therefore excluded from the photometric analysis. 

{\it Herschel/SPIRE} \citep{2010A&A...518L...1P,2010A&A...518L...3G} images at 250, 350 and 500 \micron~were obtained on 2009 December 27 with a total integration time 4408s as part of the Herschel key programme, ``MESS - Mass-loss of Evolved StarS." The reduction of the SPIRE data is described in \citet{2011A&A...526A.162G}. 
The data were processed using the standard SPIRE photometer data processing pipeline
\citep{2008SPIE.7010E..80G,2010A&A...518L...3G,2010SPIE.7731E.101D}. The procedure for calibrating the data is described in \citet{2010A&A...518L...4S}.

The {\it IRAS} imaging data have been reprocessed  using a better zodiacal light subtraction and better destriping
\citep{2005ApJS..157..302M}. The co-added 100 \micron~IRIS images have been downloaded for use in this study. 
The {\it IRAS}, {\it Spitzer} and {\it Herschel} images are shown in Figure \ref{fig7}. The new images have significantly better resolution than the {\it IRAS} 100 \micron~images. The pixel sizes are IRAS/100 (1\farcm5), MIPS/70 (5\arcsec), MIPS/160 (16\arcsec), SPIRE/250 (6\arcsec), SPIRE/350 (10\arcsec). SPIRE/500 (14\arcsec).

\section{Discussion}

R~CrB lies at high Galactic latitude (b$^{II}$=+51\degree) and so the foreground extinction is quite small, E(B-V)$\sim$0.035 mag \citep{1998ApJ...500..525S}. At maximum light, R~CrB is V=5.8 mag and B-V =0.6 mag \citep{Lawson:1990fk}. The absolute magnitude of R~CrB is estimated to be 
M$_V$=--5  mag based on the absolute magnitude/effective temperature relationship found for the Large Magellanic Cloud RCB stars \citep{2001ApJ...554..298A,Tisserand:2009fj}. So, for the analysis in this paper, we adopt a distance of 1.4 kpc. 
The most recent decline of R~CrB began in 2007 July ($\sim$JD 2454290) and the star began returning toward maximum light in the spring of 2011.
\citet{2010PASP..122..541L} point out that the recent decline of R~CrB is among the longest ever recorded.
This decline was deeper than most, bottoming out at V $\sim$15.2 mag. 

\subsection{Background Objects}

\begin{figure}
\begin{center}
\includegraphics[width=4in,angle=0]{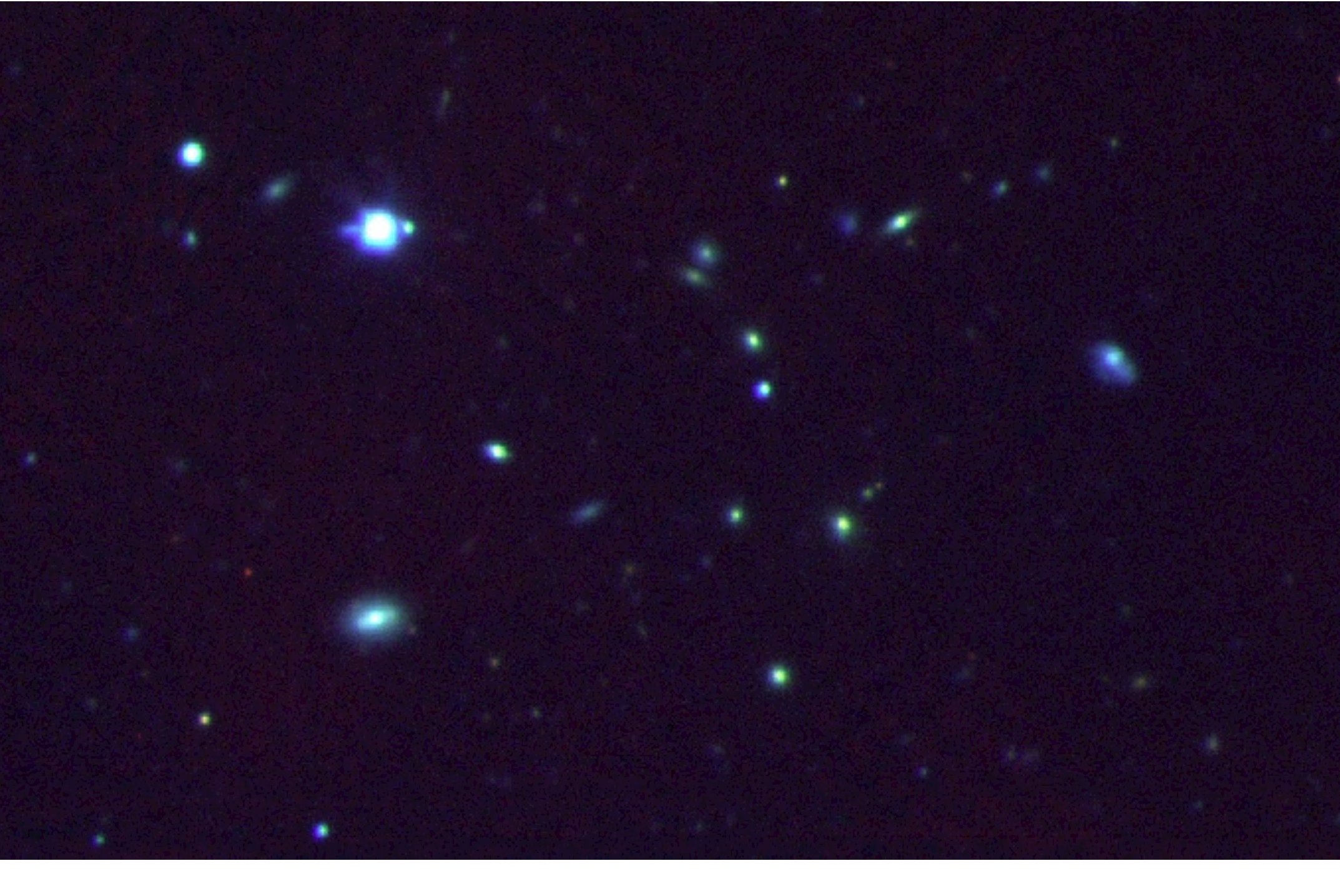}
\end{center}
\caption{{\it Gemini}/GMOS color composite image with g\arcmin~(blue), r\arcmin~(green), and z\arcmin~(red), showing the large number of galaxies near R~CrB which is the bright  object in the upper left. The field is 2\arcmin x 2\arcmin. North is up and East to the left. \label{fig2}}
\end{figure}

As shown in Figure \ref{fig1}, The {\it Gemini/GMOS} and {\it HST/WFPC2}  imaging was obtained when the star was at V$\sim$15 mag or about 9 magnitudes below maximum light. The extreme faintness of the star revealed for the first time that R~CrB  lies in front of a cluster of galaxies, shown in Figure \ref{fig2}.

A color magnitude diagram (CMD) was constructed using the {\it Gemini/GMOS} images. 
The GMOS g-band image was registered to the r-band image.  Then
SExtractor was used to obtain photometry in dual image mode, with the
r-band image used for
selecting objects.  For the g-r colors in the CMD,
photometry was obtained through 2\arcsec~diameter apertures.
The GMOS $g-r$ vs $r$ diagram, shown in Figure \ref{fig2a}, displays a fairly clear red sequence at $g-r
\sim 1.6$ and $r \sim 20$ which corresponds to a cluster of galaxies at $z
\sim 0.3$.  The sizes of the galaxies in the GMOS imaging are appropriate
for this redshift.  

R~CrB also lies in a SDSS strip obtained in 2005 when the star was bright. SDSS did not obtain spectra of any of the galaxies in Figure \ref{fig2}, but it did get spectra of about 25 galaxies within $\sim$10\arcmin~of R~CrB. Of these 14 had redshifts in the range 0.07-0.09, and 6 had redshifts in the range 0.16-0.18. 
The SDSS spectroscopy is unable to reach normal
cluster galaxies at $z \sim 0.3$ and the SDSS redshifts are likely to be
those of field galaxies in the foreground of the cluster.  

\citet{2010PASP..122..541L} were the first to point out that there is a faint red star $\sim$3\arcsec~west of R~CrB. They find that the star has a magnitude of $V$=20.78$\pm$0.06 and $B-V$ = 0.99$\pm$0.08. It is also clearly visible in the {\it HST}, and {\it Gemini} images taken during the R~CrB minimum in Figures \ref{fig3} and \ref{fig5}. 
If the star is a physical companion, it would be at a projected
distance of about 4000 AU from R~CrB, and its absolute magnitude would be
$\sim$+10 mag. This would imply that the star is a main sequence M
star, however the observed $B-V$ colors are not consistent with such a star.
 On the other hand, if the faint companion to R~CrB is a K2 V star, which is more consistent with its colors ($B-V$ = 0.91 and
 $M_V$=+6.4 mag \citep{2000asqu.book.....C}), then it is at a distance
 of $\sim$7 kpc.  In short, the companion star is likely to be a background K dwarf. 
 The companion star to R~CrB is not visible in the {\it HST/WFPC2} F469N images from 1996. The WFPC2 ETC, assuming a 20th magnitude K0 star, estimates that the ratio of the count rates in the three filters will be F469N/F555W/F814W = 0.04/5.7/6.7 electrons pixel$^{-1}$ s$^{-1}$. So it is likely that the star was too faint to be seen in the F469N images.

\subsection{Cometary Knots Close to R~CrB}

\begin{figure}
\begin{center}$
\begin{array}{ccc}
\includegraphics[width=2.2in,angle=0]{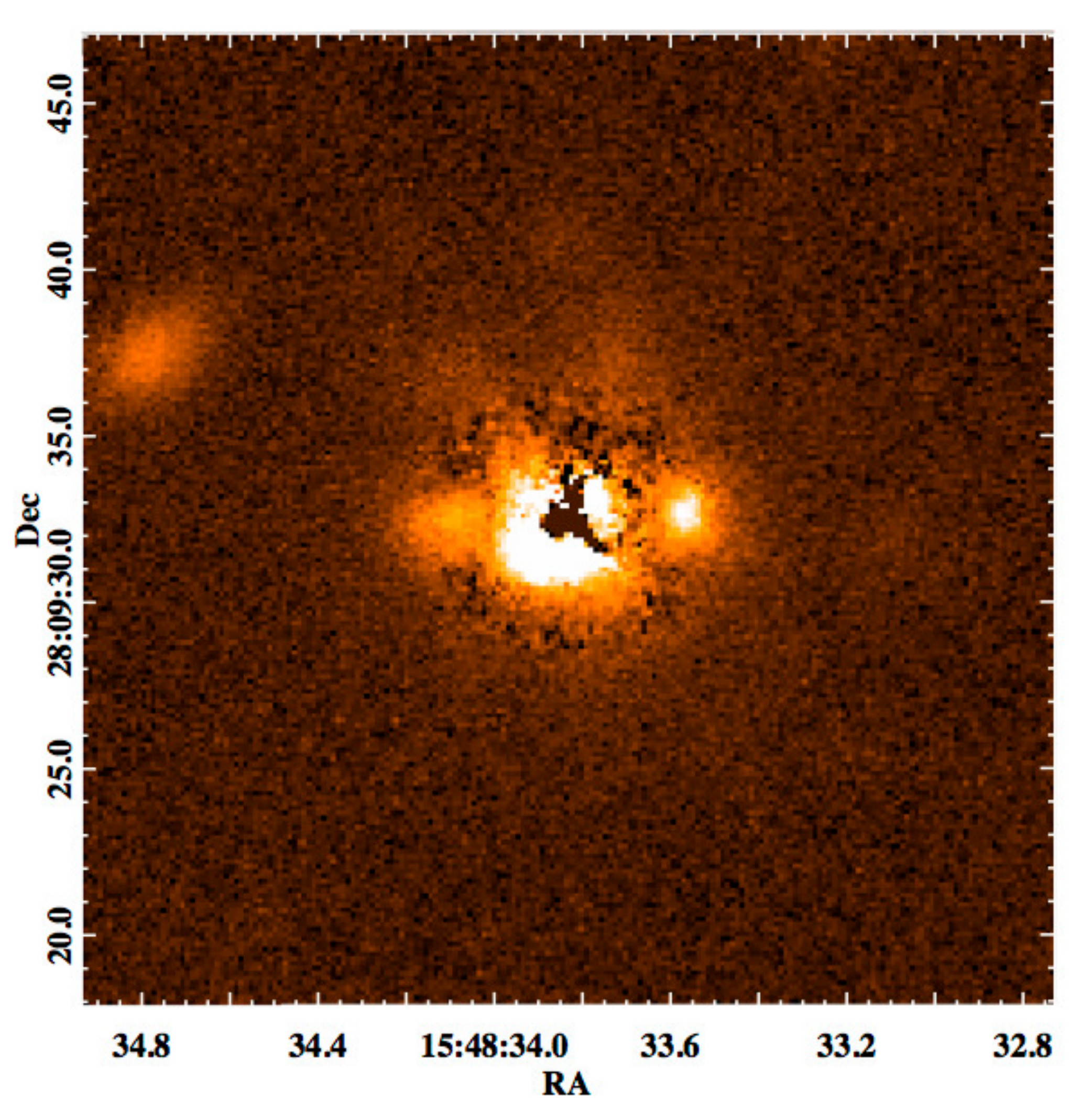}&
\includegraphics[width=2.2in,angle=0]{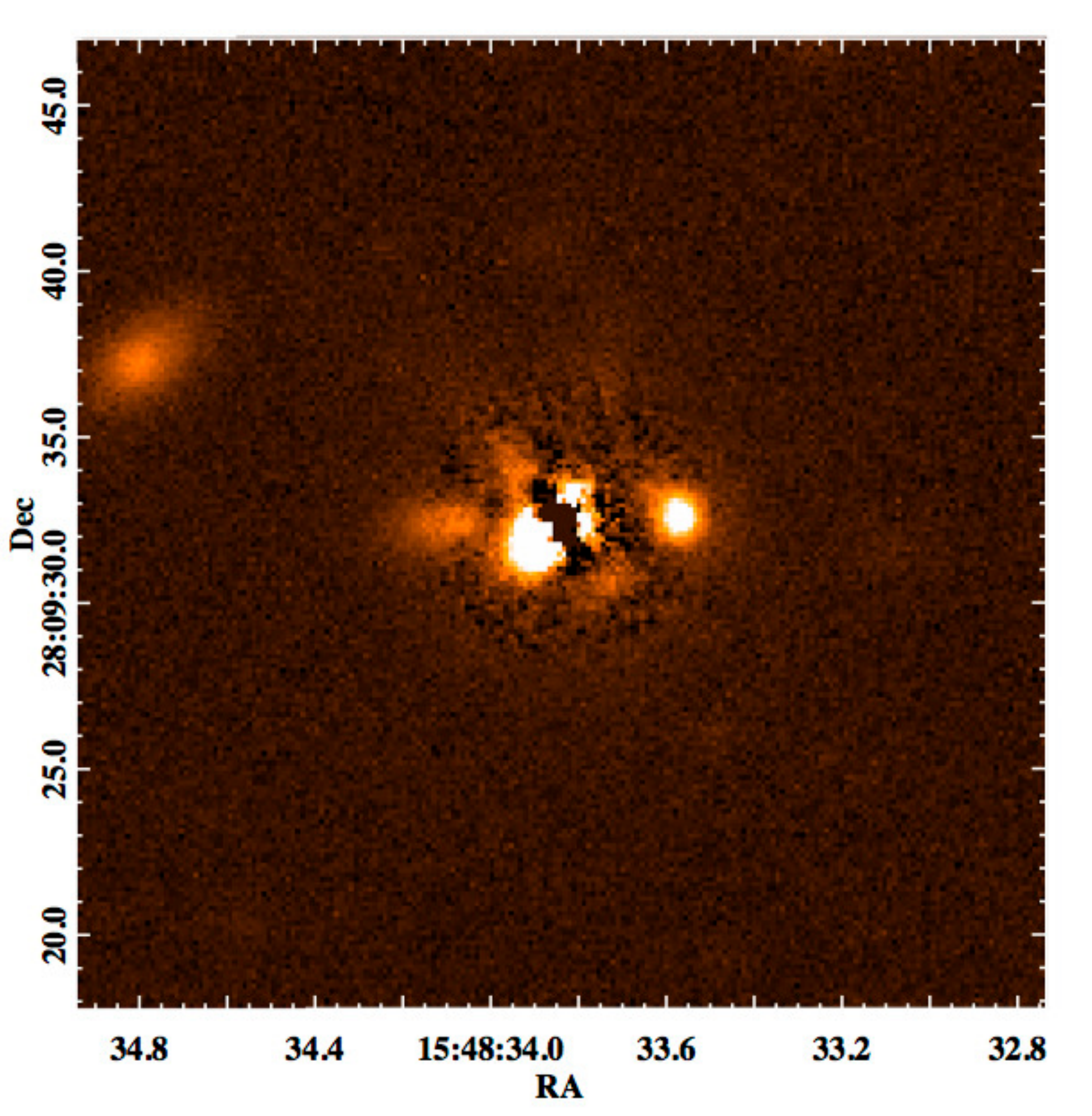}&
\includegraphics[width=2.2in,angle=0]{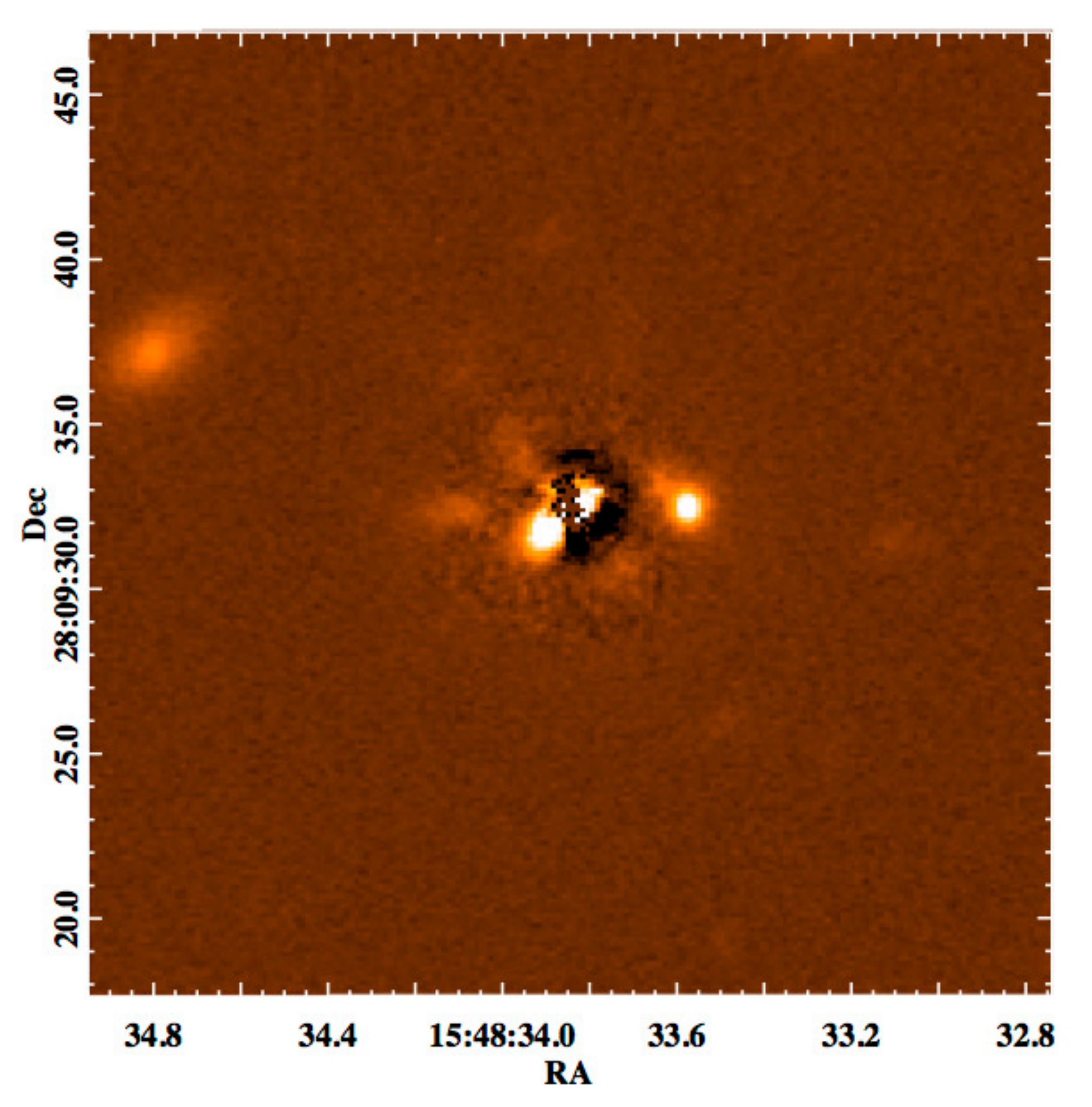}\\
\includegraphics[width=2.2in,angle=0]{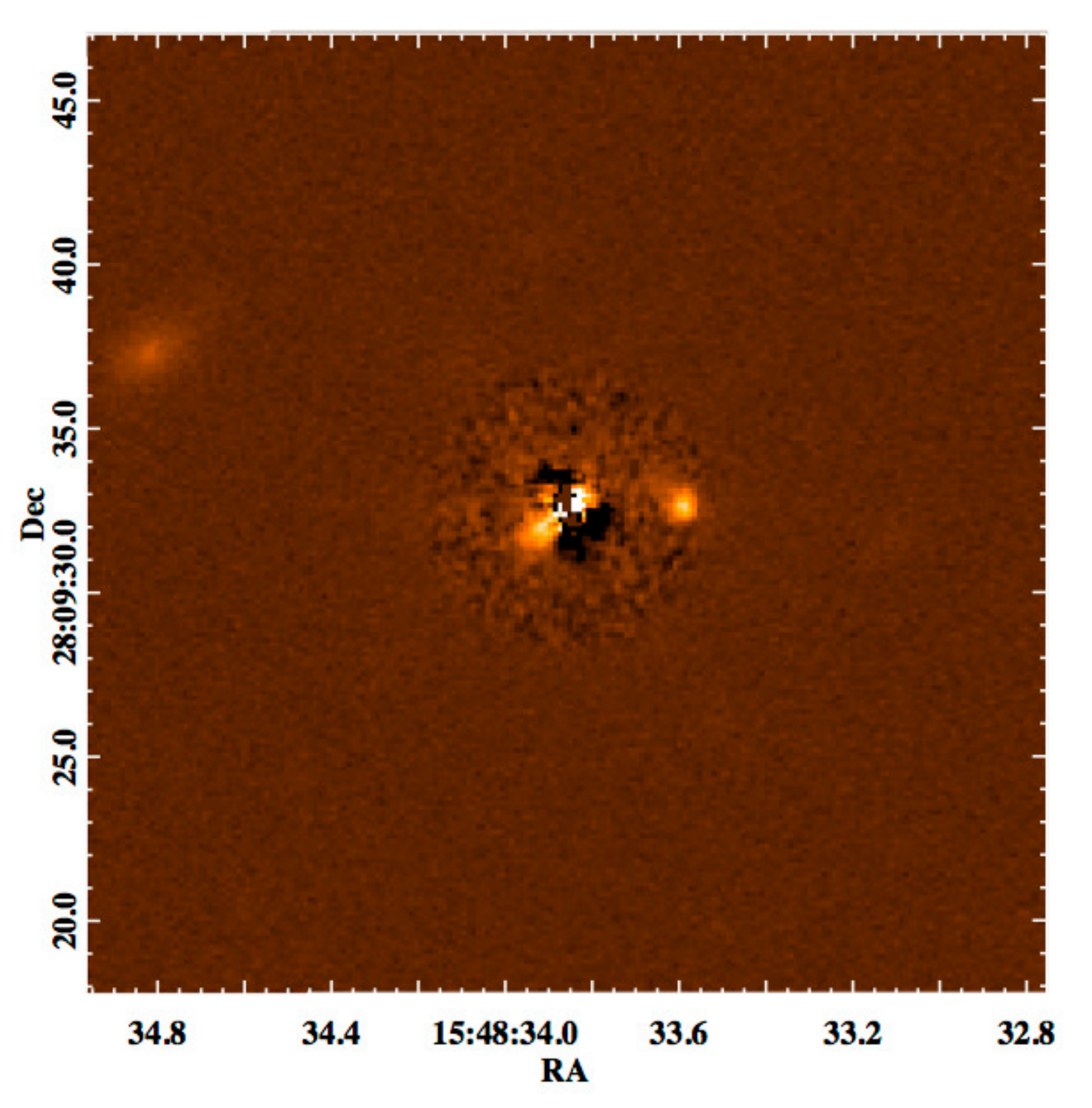}&
\includegraphics[width=2.2in,angle=0]{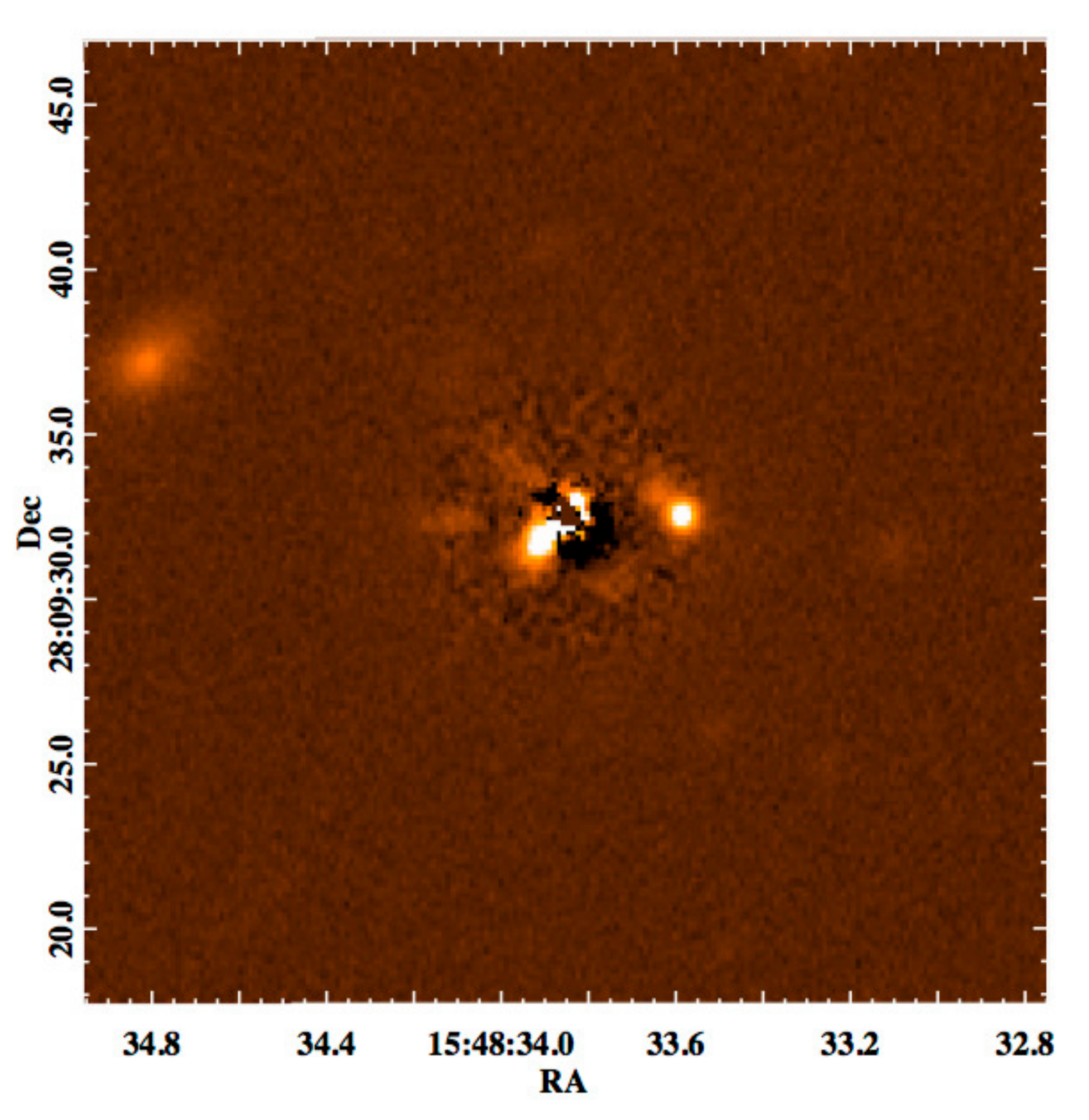}
\end{array}$
\end{center}
\caption{{\it Gemini}/GMOS g\arcmin, r\arcmin, i\arcmin, z\arcmin, and CaT images. A stellar PSF has been subtracted from the position of R~CrB for each image. There is a faint star 3\arcsec~west of R~CrB and a galaxy about 15\arcsec~northeast of R~CrB. North is up and East to the left. \label{fig3}}
\end{figure}

\begin{figure}
\begin{center}
\includegraphics[width=3in,angle=0]{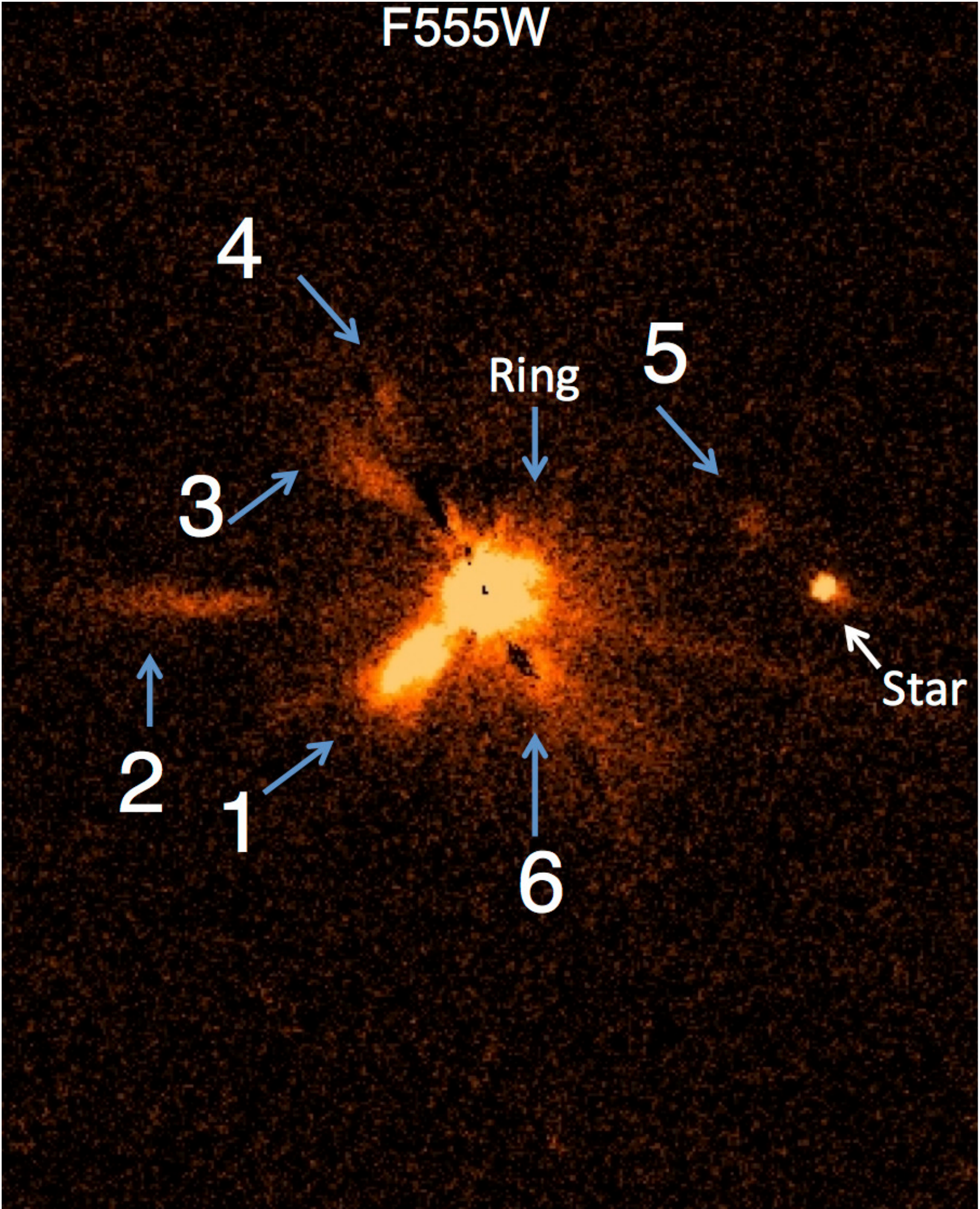}
\includegraphics[width=2.76in,angle=0]{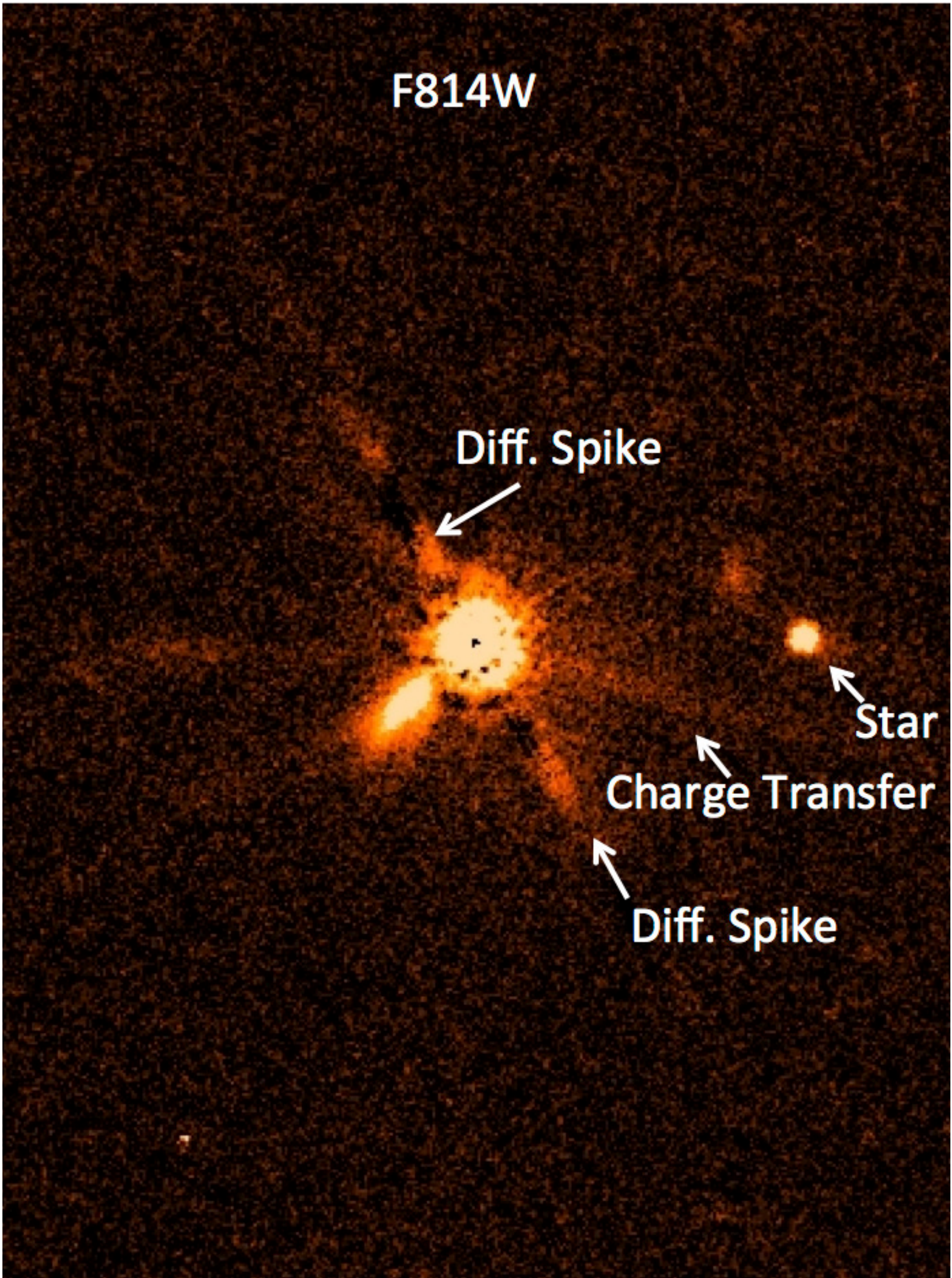}

\end{center}
\caption{{\it HST}/WFPC2 F555W (left) and F814W (right) images with a PSF removed from the position of R~CrB. White arrows indicate instrumental features and the companion star. Blue arrows indicate real nebular structures around R~CrB. The angular distance between R~CrB, which lies in the center of the images, and the faint star to the west is $\sim$3\arcsec. The cometary knots are numbered.\label{fig5}}
\end{figure}

\begin{figure}
\begin{center}
\includegraphics[width=6in,angle=0]{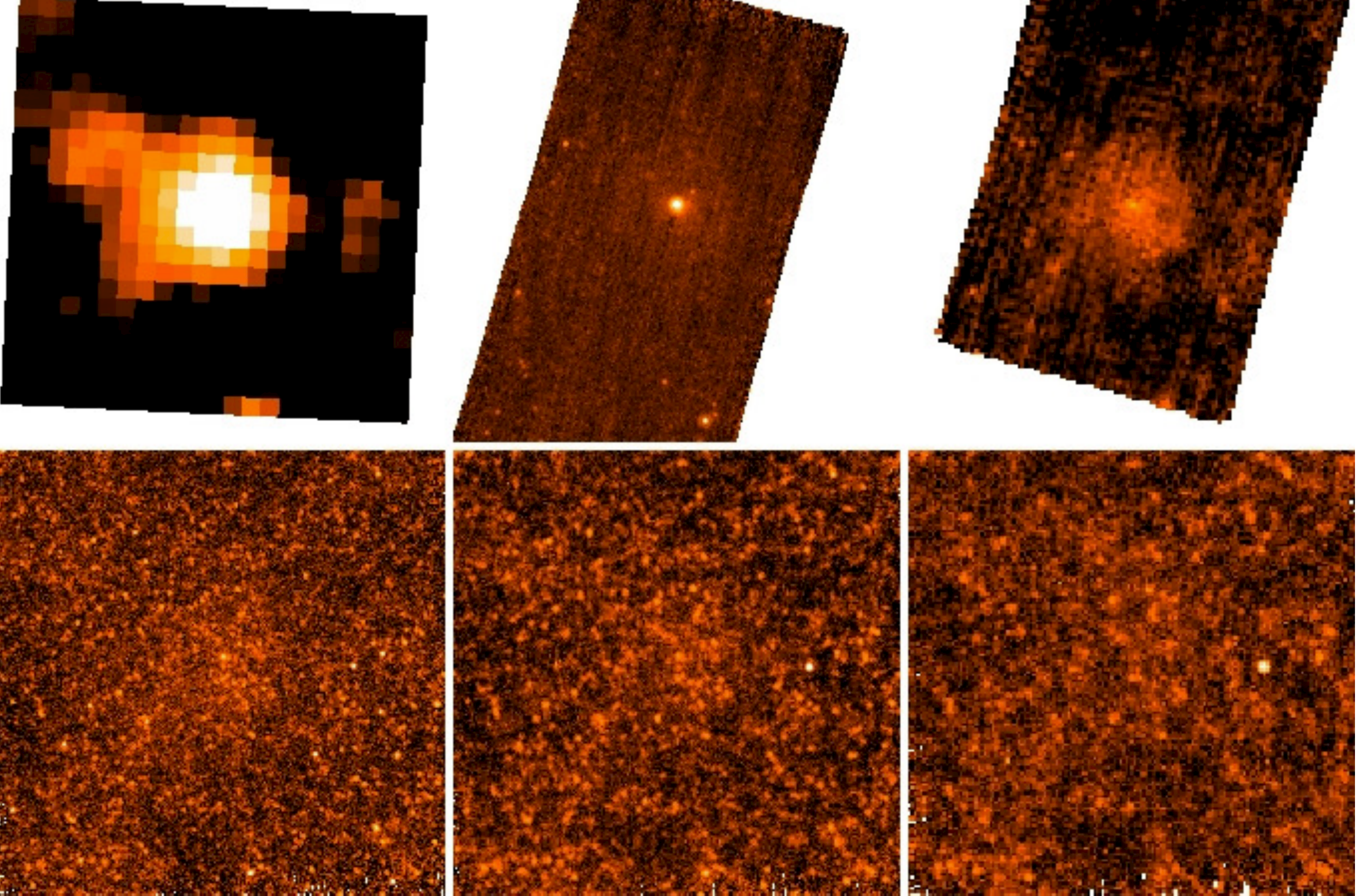}
\end{center}
\caption{Far-IR images centered on R~CrB showing the large shell with a radius of 10\arcmin. Upper: {\it IRAS/IRIS} 100 \micron, {\it Spitzer/MIPS} 70, and 160 \micron, Lower: {\it Herschel/SPIRE} 250, 350, and 500 \micron. The field is 35\arcmin~x 35\arcmin. \label{fig7}}
\end{figure}

\begin{figure}
\begin{center}
\includegraphics[width=5in,angle=0]{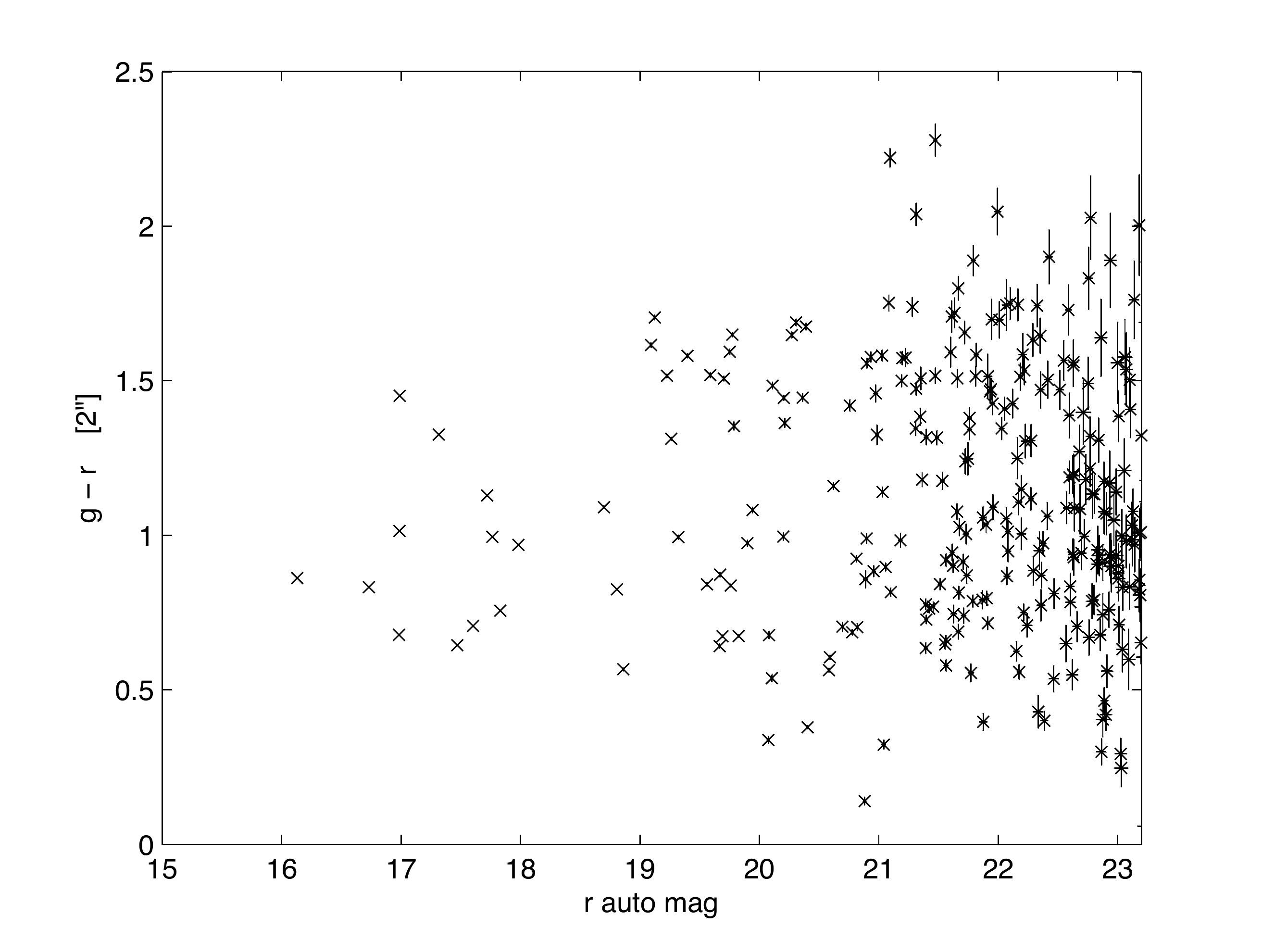}
\end{center}
\caption{The CMD based on the {\it Gemini/GMOS} images. There is a clear red
sequence at g-r = 1.6 which suggests a cluster at z$\sim$0.3. The clustering at g-r$\sim$1 is probably M stars and foreground galaxies. \label{fig2a}}
\end{figure}

The {\it Gemini} and {\it HST} fields centered on R~CrB are shown in Figures \ref{fig3} and \ref{fig5}. A stellar PSF has been subtracted from the position of R~CrB for each image. The {\it Gemini} and {\it HST} images were obtained only a month apart and the same structures can be seen in both. The  structures are labeled in Figure \ref{fig5}, showing real nebulosity as well as instrumental artifacts such as diffraction spikes and charge transfer effects. The real structures appear in multiple filters in both GMOS and WFPC2 images so are unlikely to be artifacts. 

R~CrB was 9 magnitudes fainter than at maximum light during the decline. The dust lying in front of the star is extremely optically thick so 
at this time we are seeing essentially zero directly transmitted light. 
When the star itself is completely obscured, all that remains is a fairly constant scattered-light residual
that is likely to be from photons that have escaped around the edge of the obscuring cloud by scattering off dust lying beside or behind the star. This scattered light may be responsible for the typical flat-bottomed declines often seen in RCB star lightcurves. 
When the stellar photosphere is obscured in deep declines, RCB stars show emission lines in their spectra. Spectra were obtained of R CrB in 2008 when it was deep in the present decline showing emission lines including the Na I D lines and the IR Ca II triplet \citep{Kameswara-Rao:2010lr}. 
These spectra show that the emission lines account for no more than 5-10\% of the flux in the V band. 
R~CrB, itself, appears to be marginally resolved in the WFPC2 images. There is a diffuse, nearly circular ``halo" around the star. This may be the dust near the star that is scattering the light around the obscuring cloud. 
\citet{jeffers2011} also suggest the presence of this halo based on the brightness of R~CrB at the bottom of its decline. 
The blobs visible in Figures \ref{fig3} and \ref{fig5} are also seen in scattered light from R~CrB. 

\begin{figure}
\begin{center}
\includegraphics[width=6in,angle=0]{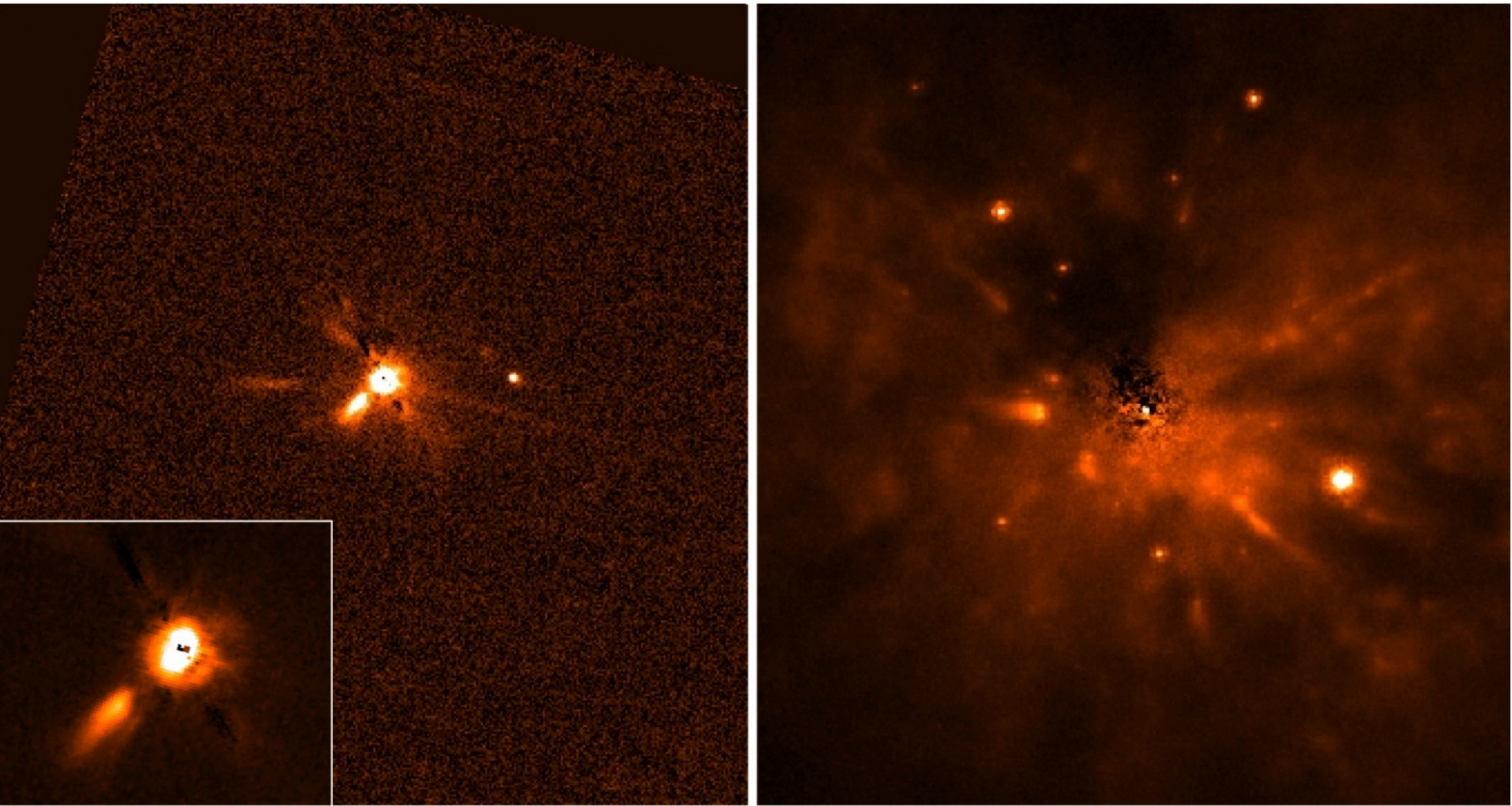}
\end{center}
\caption{{\it HST/WFPC2} F555W image (17\arcsec~x 17\arcsec)  of R~CrB (left) and WFPC2 606W image (12\arcsec~x 12\arcsec) of UW Cen (right). The inset shows the cometary globule (\#1) to the lower left of R~CrB. \label{fig5a}}
\end{figure}

\begin{figure}
\begin{center}
\includegraphics[width=4in,angle=0]{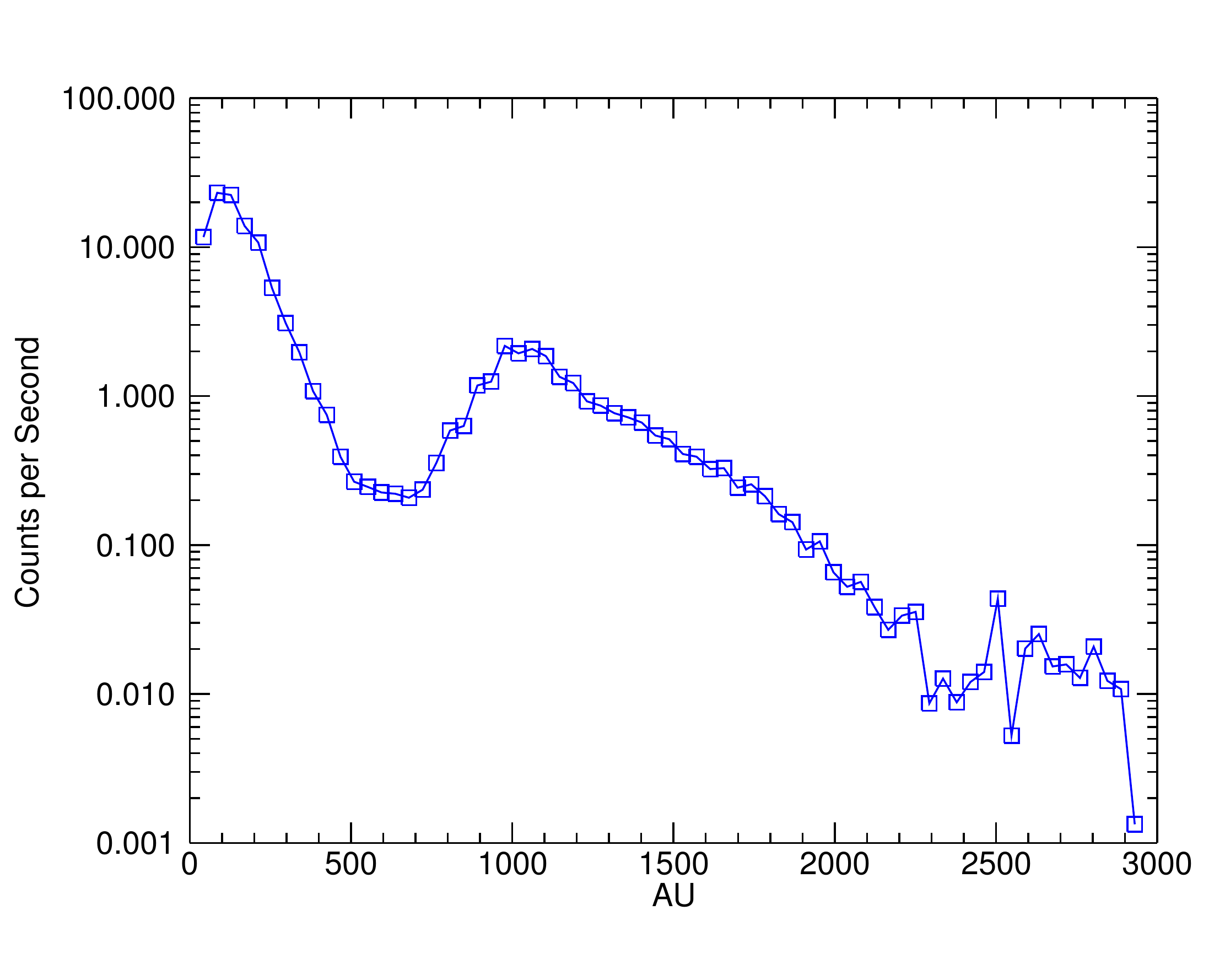}
\end{center}
\caption{A cut from the center of the PSF-subtracted through the large cometary knot  (\#1) in the HST/WFPC2 F555W image. The cometary globule lies at 1000-2000 AU from the star. \label{fig5b}}
\end{figure}

Several of the structures around R~CrB appear as jet-like blobs pointing radially away from the star. This could either be a lighting effect created by the central star 
shining through holes in the dust clouds and illuminating different portions of the
circumstellar dust shell, or the blobs could be ``cometary" knots. The RCB star, UW Cen, shows both illumination structures and cometary structures \citep[][Clayton et al. 2011, in preparation]{1999ApJ...517L.143C}. Figure \ref{fig5a} shows a comparison of the blobs seen in R~CrB and cometary globules seen in UW Cen. In particular, the globule to the SE of R~CrB shows the typical ``tadpole" shape similar to those seen in the Helix nebula \citep{2009ApJ...700.1067M}. It also shows a sharp edge on the side toward the star with the brightness fading away from the star, as shown in Figure \ref{fig5b}. 
The largest knot (\#1 in Figure  \ref{fig5a}) is $\sim$1000 AU across. This is larger than the knots in the Helix which are 100-300 AU in size with masses of 10$^{-5}$M$_{\sun}$ \citep{1996AJ....111.1630O}. 
The simplest explanation for these cometary knots is that they are caused by a stellar wind streaming past a density enhancement moving slowly away from the star \citep{2006ApJ...646L..61G}. 
There is evidence for both low (10-20 km s$^{-1}$) and high (200-400 km s$^{-1}$) velocity gas streaming away from RCB stars \citep{1992ApJ...397..652C,2003ApJ...595..412C,2011arXiv1107.1185G}.
Simulations show that when a slow wind is overtaken by a fast wind, the gas-gas interaction could create knot-like shapes \citep{2005MNRAS.361.1077P}. Recent investigations using a hydrodynamics code including dust shows that small dust grains ($\lesssim$0.045 \micron) will follow the motion of the gas closely \citep{2011ApJ...734L..26V}. 
The UW Cen nebula, about 15\arcsec~across, resembles a fossil PN shell where the gas is neutral and the shell is seen in light scattered from dust  \citep{1999ApJ...517L.143C}. UW Cen is at a distance of $\sim$5.5 kpc so if it were at the distance of R~CrB, its nebula would be about 1\arcmin~in size. 
As described below, the diffuse nebulosity around R~CrB extends at least 10\arcmin~from the star but no blobs or density concentrations are seen more than a few arcseconds from the star. 

\citet{jeffers2011} report imaging polarimetry of the cometary knot SE of R~CrB (\#1 in Figure \ref{fig5a}). Combining those data with the WFPC2 images presented here, they infer that the knot is at a scattering angle of 105\degree~and has a dust mass of $\sim$10$^{-6}$ M$_{\sun}$. It lies about 1\arcsec~from R~CrB which is about 1400 AU from the star. At velocities of 20 and 200 km s$^{-1}$, the age of the blob would be 300 or 30 yr, respectively. 

To estimate the mass of dust present in the knots, we make the
following simplifying assumptions: (1) the illuminating flux from R
CrB is unobscured (i.e.\ equal to the maximum values given in Table
2), (2) the dust in the knot is optically thin, such that all grains
are illuminated and no grains are shadowed, and (3) that the grains
can be described by the neutral PAH and graphitic solid properties
from \citet{2001ApJ...554..778L}.  Following \citet{1986ApJ...308..225C}, the flux scattered off
one dust grain of radius $a$ at a distance $r$ from a source with flux
$F(\lambda)$ is
\begin{equation}
dF_{sca}(\lambda,r,a) = \frac{Q_{sca}(\lambda,a)\pi a^2
\Phi(\lambda,a,\theta)F(\lambda)}{4\pi r^2}
\end{equation}
where $Q_{sca}$ is the scattering efficiency, $\Phi$ is the
\citet{1941ApJ....93...70H} phase function, and $\theta$ is the scattering angle.  If
one assumes the dust grains have uniform size, the mass of
grains ($M_{gr}$) in a knot that produces a total scattered flux ($F_{sca}$) is
\begin{equation}
M_{gr} = \frac{16\pi^2\rho a r^2 F_{sca}}
 {3F(\lambda)Q_{sca}(\lambda,a) \Phi(\lambda,a,\theta)}
\end{equation}
where $\rho=2.24$~g~cm$^{-3}$ is the density of carbonaceous dust.
If one instead adopts an MRN size distribution \citep{1977ApJ...217..425M} such that
the number density of grains $n_{gr}\propto a^{-3.5}$ then
\begin{equation}
M_{gr} = \frac{4r^2F_{sca}}
{F(\lambda)\int 0.224 \pi a^{-1.5}
 Q_{sca}(\lambda,a)\Phi(\lambda,a,\theta)da}
\end{equation}
where the numeric constant is the normalization factor from
\citet{2001ApJ...548..296W}, and assuming a gas-to-dust ratio of 100.    Finally, if the
dust is assumed to lie in a spherical shell of inner and outer radii
$R_{in}$ and $R_{out}$ then $M_{gr}$ must be integrated over the
volume of the shell.  Since
\begin{equation}
\int_0^\pi \Phi(\lambda,a,\theta) \sin{\theta} d\theta = 2
\end{equation}
the total grain mass for a single-sized model equals
\begin{equation}
M_{gr} = \frac{16\pi\rho a F_{sca}}{9 F(\lambda)Q_{sca}(\lambda,a)}
\frac{R_{out}^3 - R_{in}^3}{R_{out}-R_{in}}.
\end{equation}

Of the knots identified in Fig.\ 4, numbers 4 and 6 are coincident
with diffraction spikes, and we refrain from estimating their dust
masses.  For the others, the total flux from each knot was measured
through manually-sized elliptical apertures.  These values are
reported in Table 3.  For each knot, models with single grain
sizes of 0.001, 0.01, 0.1 and 1.0 $\mu$m, as well as full MRN
distributions, were tested for scattering angles from 10\degr to 170\degr~(in
increments of 10\degr) to find which combinations of grain sizes and
scattering angles yielded equivalent dust masses for both the F555W
and F814W photometry.  The best-fitting models, including the grain
sizes, scattering angle, resulting clump distance from R~CrB and
estimated dust mass, are also listed in Table 3.  
The model dust masses are 10$^{-10}$ to 10$^{-7}$ M$_{\sun}$.
Note that
if a model is not listed, it failed to reproduce the photometry in
both filters.

Knot 3 has very low flux in the F814W filter, resulting in a large
uncertainty with the flux between 4 and 6 $\times 10^{-18}$
erg~cm$^{-2}$~s$^{-1}$~\AA$^{-1}$.  For this reason, results are shown
for the upper and lower flux range. Since small grains scatter
isotropically, there were no preferrable scattering angles for this
knot using the smaller single grain-size models, making a distance
estimate from R~CrB impossible.  Owing to the very
bright F814W flux from knot 5, we find there is no dust-scattering
model that adequately reproduces the fluxes in both filters.

To estimate the dust mass producing the halo (i.e., the
slightly-resolved profile around R~CrB), we assume that all of its
flux is light scattered by a thin circumstellar shell.  We find that a
limb-brightened shell of optically-thin dust can be up to 0\farcs3 in
radius and still produce a profile consistent with that of R~CrB in
the 2009 HST images.  The flux of the halo, as well as the
best-fitting dust model that produces this scattered light, is listed
at the bottom of Table 3.

Note that our first modeling assumption implies that the masses
derived above are lower limits, since more dust is required if the
incident flux is lower than that of R~CrB at maximum light.  As a
quick check of the second assumption, we assume the knots are
ellipsoids with the same axes as those used to measure the fluxes.  In
knot 1, there are  $8\times 10^{38}$ grains of radius 0.1 $\mu$m,
corresponding to a number density of $n_{gr}=8\times 10^{-10}$
cm$^{-3}$.  The optical depth $\tau=n_{gr} L (Q_{sca} + Q_{abs}) \pi
a^2$ where $L$ is the line-of-sight distance through the dust and
$Q_{abs}$ is the absorption efficiency.  The proposed ellipsoid has a
depth around $10^{16}$ cm, yielding an approximate $V$-band optical depth
of $8\times 10^{-3}$, which is sufficiently thin to warrant the
assumption that all grains are illuminated by the star. See \S3.3 for
a justification of the last assumption, namely, that the dust is purely
carbonaceous.

\subsection{The Large Far-IR Dust Shell}

\begin{figure}
\begin{center}
\includegraphics[width=4in,angle=0]{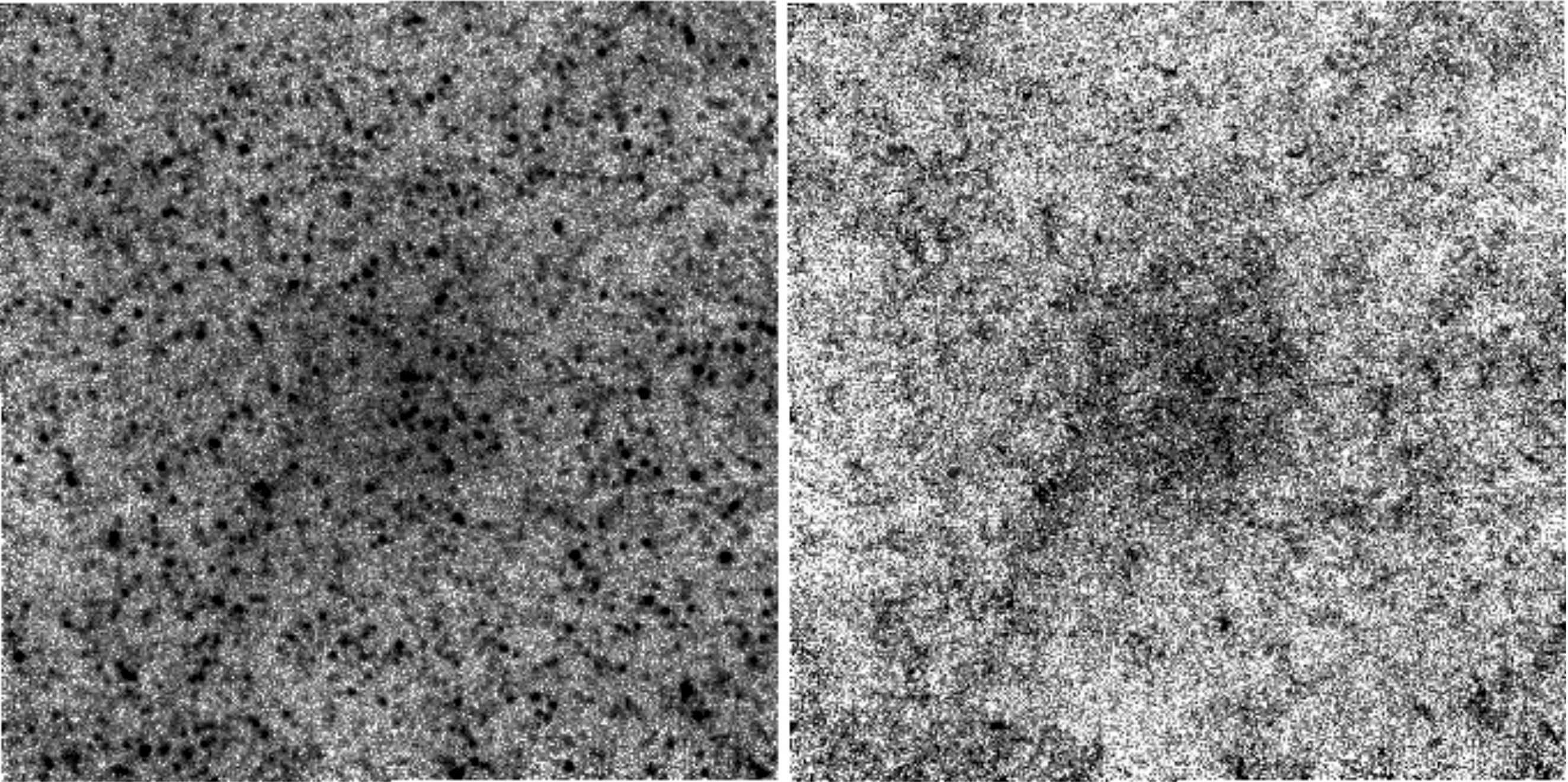}
\end{center}
\caption{Left: {\it Herschel}/SPIRE 250 \micron~image centered on R~CrB. Right: Same image with points sources subtracted to show diffuse IR emission is present. The field is 34\arcmin~x 34\arcmin.  \label{fig9}}
\end{figure}

\begin{figure}
\begin{center}
\includegraphics[width=4in,angle=0]{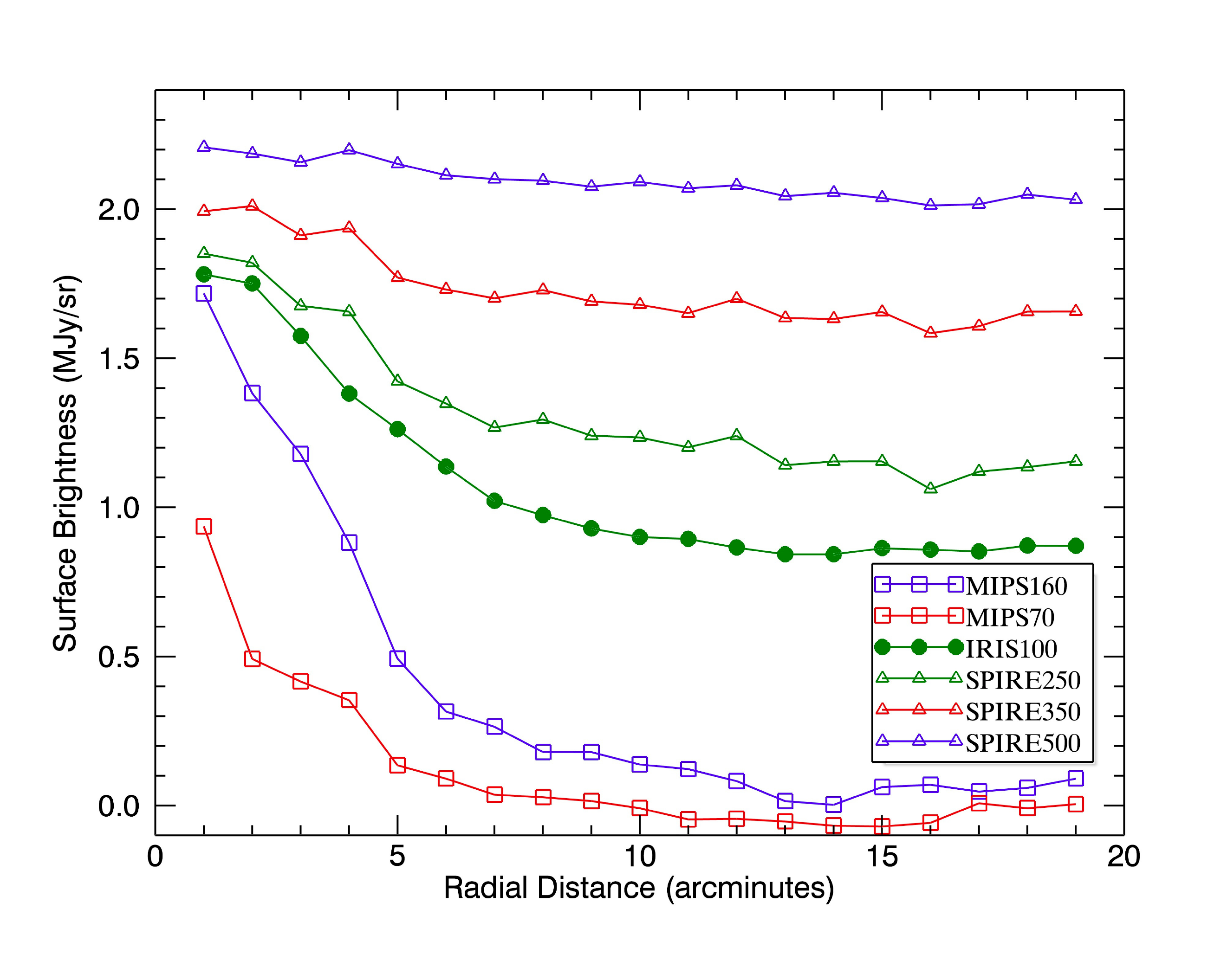}
\end{center}
\caption{Surface brightness of diffuse emission surrounding R~CrB in one arcminute rings. The curves have been shifted vertically for clarity. \label{fig10} }
\end{figure}

R~CrB has an {\it IRAS} shell with a radius of 10\arcmin, 4 pc at a distance of 1.4 kpc \citep{Gillett:1986cr}. The new MIPS images at 70 and 160 \micron, shown in Figure \ref{fig7} along with the {\it IRAS} 100 \micron~image, confirm the presence of a large dust shell. It is also visible in the SPIRE images at 250 and 350 \micron. It is not clearly visible at 500 \micron. The new MIPS and SPIRE images have significantly better resolution than {\it IRAS}. 
These new images imply that the shell is spherical in shape and that the component to the northeast of the nebula in the {\it IRAS} 100 \micron~image is detached from the R CrB shell in the MIPS and SPIRE images. 
The SPIRE images, in particular,  are full of point sources, which are background galaxies. 
To determine how much of the nebulosity seen in the IRAS, MIPS and SPIRE images might be due to blended emission from background galaxies, we used Sextractor to identify and remove point sources from the SPIRE 250 \micron~image. The result is shown in Figure \ref{fig9}. This clearly shows that the dust nebulosity around R~CrB is not due to a concentration of galaxies in a cluster behind the star. 
There is significant diffuse far-IR emission, centered on R~CrB. 
To get a clearer idea 
of the extent of the nebulosity at different IR wavelengths, the average brightness in 1\arcmin~concentric rings from R~CrB was calculated and plotted in Figure \ref{fig10}. These new data show that the shell is possibly even larger than that seen by {\it IRAS}. 
The MIPS 70 \micron~image also has a central elongated structure running almost east-west that has a diameter of $\sim$3\arcmin. 

 


At the assumed distance of R~CrB, the very extended IR shell extends about 8 $\times$ 10$^5$ AU from the star.
The dust mass and total mass of this shell have previously been estimated to be $\sim$10$^{-3}$--10$^{-2}$ M$_{\sun}$ and $\sim$0.3 -- 3 M$_{\sun}$, respectively \citep{Gillett:1986cr}. They estimate that the dust temperature in the
shell is 25-30 K. 
\citet{1986MNRAS.222..357K} find a hot component close to the star with a dust mass of 5 $\times$ 10$^{-7}$
M$_{\sun}$ and a cool component with a dust mass of 10$^{-2}$ - 10$^{-3}$ M$_{\sun}$. 

\begin{figure}
\begin{center}
\includegraphics[width=5in]{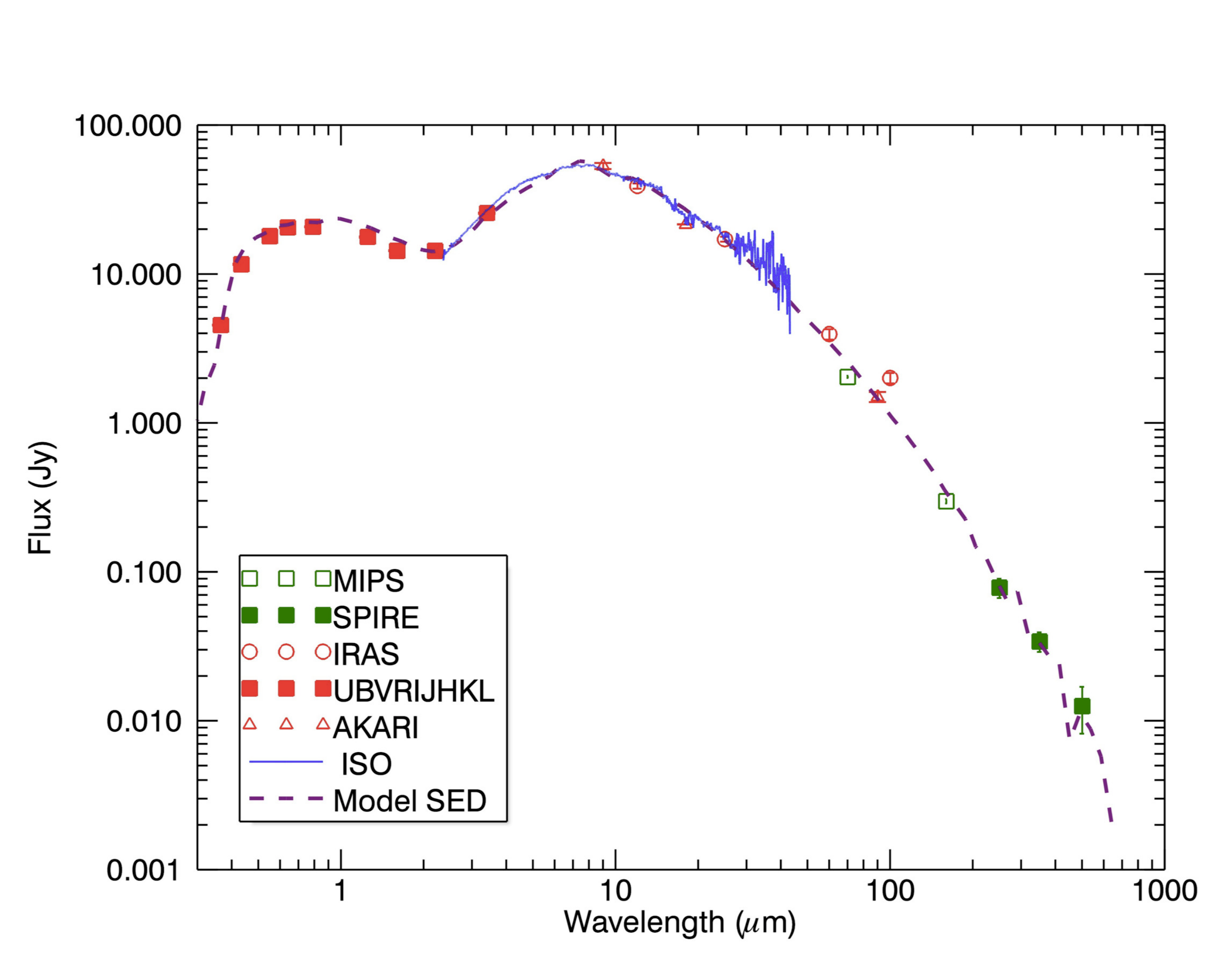}
\end{center}
\caption{Maximum light SED for R~CrB. Filled red squares: UBVRIJHKLM, open red circles: IRAS (12, 25, 60, 100 \micron), open green squares: MIPS (70, 160 \micron), filled green squares: SPIRE (250, 350, 500 \micron), open red triangles: AKARI 9, 18, 90 \micron), blue line: {\it ISO} spectrum. The dust associated with R~CrB was clearly detected in all the SPIRE bands. Also plotted is the best fitting radiative transfer model to the SED (purple dashed line).  \label{fig11}}
\end{figure}

Photometry available from the literature for R~CrB has been combined with new photometry using the MIPS and SPIRE images.  The data, summarized in Table 2 and plotted in Figure  \ref{fig11},  consist of visible photometry at maximum light, JHKL, as well as {\it IRAS} photometry  \citep{Walker:1985rr, 1986MNRAS.222..357K}. 
AKARI/IRC and FIS photometry are also available \citep{2007PASJ...59S.369M,2010A&A...514A...1I}.
Fluxes were extracted from  SPIRE (250, 350, 500 \micron) using Starfinder \citep{2000A&AS..147..335D}. 
Starfinder produced PSF fitted 
photometry using PSFs which were downloaded from the Herschel Science
Center.  No aperture corrections were applied since the PSFs used were of sufficiently
large diameter. Aperture photometry was also performed as a test of the Starfinder results. 
The photometric uncertainties quoted in Table 2 for SPIRE are 15\%, or the Starfinder uncertainties plus calibration errors of 5\% whichever is higher \citep{2010A&A...518L...4S}. 

We have modeled the SED of R~CrB using a Monte Carlo radiative transfer code which includes nonisotropic scattering, polarization, and thermal emission from dust in a spherical-polar grid \citep{2003ApJ...598.1079W,2003ApJ...591.1049W,2006ApJS..167..256R}. All of the models were done with amorphous carbon dust and an MRN size distribution \citep{1977ApJ...217..425M}. We have assumed a mass gas-to-dust ratio of 100.  The best fit to the R~CrB SED is shown in Figure \ref{fig11}.  A good fit was obtained to both the hotter dust centered on the star, as well as the cooler, large diffuse nebula. Heating by the interstellar radiation field is included. 
The best fitting model is a small disk surrounded by a large diffuse envelope. 
The disk has inner and outer radii of 42 and 150 AU, and the envelope has inner and outer radii of 1.3 and 8.8 $\times$ 10$^5$ AU. 
The envelope has a bipolar cavity carved out with an opening angle of 30\degree.
The A$_V$ through the envelope is quite low, less than 0.01 mag over a wide range of viewing angles (0\degree-75\degree). 
The disk dust  mass is 3.5 $\times$ 10$^{-6}$ M$_{\sun}$ and a total mass of 3.5 $\times$ 10$^{-4}$ M$_{\sun}$. 
Assuming a density profile of r$^{-2}$ which is appropriate for dust accelerated away from the star radially by radiation pressure, the dust mass of the outer envelope is  2 $\times$ 10$^{-2}$ M$_{\sun}$ and  the total mass (dust + gas) is 2 M$_{\sun}$. The gas may be dragged along with the dust or it may become decoupled.

\subsection{The Evolution of R~CrB and its Dust}

The generally accepted model for dust formation in the RCB stars, the condensation of carbon dust along the line of sight,
was suggested over 70 years ago \citep{1935AN....254..151L,1939ApJ....90..294O,1996PASP..108..225C}. 
The evolution of RCB star spectra and lightcurves during declines is consistent with dust that forms close to the stellar atmosphere and then is accelerated to hundreds of km s$^{-1}$~by radiation pressure \citep{1992ApJ...397..652C,1993ASPC...45..115W}.
If the dust forms near the star, it will accelerate very quickly, possibly reaching 200 km s$^{-1}$ in 30 days. 
Other evidence supports a much slower expansion \citep[e.g.,][and references therein]{2011arXiv1107.1185G}.
The observed timescales for 
RCB dust formation fit in well with those calculated by carbon chemistry models \citep{1986ASSL..128..151F,1996A&A...313..217W,Kameswara-Rao:2010lr}.

The dust forming around R~CrB is not in a complete shell but rather in small ``puffs" \citep{1996PASP..108..225C}. When a puff forms along the line of sight to the star we see a decline.
Estimates of the covering factor of the clouds around RCB stars during declines, from both extinction studies and IR re-emission of stellar radiation, indicate that f $<$ 0.5 \citep{Feast:1997lr,1999ApJ...517L.143C,1984ApJ...280..228H,1998ApJ...501..813H}. 
The typical puff causing a decline is thought to have a dust mass of $\sim$10$^{-8}$ M$_{\sun}$ \citep{1986ASSL..128..151F,1992ApJ...397..652C}.
There is strong evidence that the puffs form in or near the surface of the RCB star due to density and temperature perturbations caused by stellar pulsations \citep{1996A&A...313..217W,2007MNRAS.375..301C}. The RCB stars show regular or semi-regular pulsation periods in the 40-100 d range \citep{Lawson:1990fk}. R~CrB, itself, does not have one regular period but has shown periods of 40 and 51 d \citep{1993MNRAS.265..899F}. 

For this recent deep decline, with A$_V$=9 mag, assuming the dust forms at 2 R$_{\star}$ (R$_{\star}$=85 R$_{\sun}$), and a puff subtends a fractional solid angle of 0.05, then the dust mass will be $\sim$10$^{-8}$ M$_{\sun}$. The puff would be accelerated quickly by radiation pressure and would dissipate rapidly so dust must be formed continually by R~CrB to maintain itself in a deep decline for 4 years. If a puff is formed during each pulsation period of $\sim$40 d, R~CrB would be forming about 10$^{-7}$ M$_{\sun}$ of dust per year. Assuming a gas-to-dust ratio of 100 \citep{2003dge..conf.....W}, the total mass loss per year is 10$^{-5}$ M$_{\sun}$. 
The estimated masses of the puffs causing the declines and the cometary knots is similar. The puffs cause declines when they form directly in our line of sight and may be seen as cometary knots when they form to the side of or behind R~CrB. 

Little is known about the lifetime of the
RCB phase.  We have a lower limit from the fact that R~CrB itself was
discovered to have large brightness variations 200 years ago \citep{1797RSPT...87..133P}.
Assuming a velocity of 20 km s$^{-1}$ then the large diffuse dust shell around R~CrB  would take  $10^5$ yr to form. If the mass-loss was more like the high-velocity winds seen in R~CrB today ($\sim$200 km s$^{-1}$) then the shell would be about an order of magnitude younger \citep{2003ApJ...595..412C}. 
If R~CrB is the result of a final-helium-shell flash rather than a white-dwarf merger, then the size and timescales would be consistent with the nebulosity, now seen in far-IR emission, being a fossil PN shell. The nebulosity including cometary knots, seen around R~CrB and UW Cen, shown in  Figure \ref{fig5a}, looks very much like a PN shell \citep[][Clayton et al. 2011, in preparation]{1999ApJ...517L.143C}. 

Planetary Nebulae are a very late phase in the evolution of low and intermediate mass (1-8 M$_{\sun}$) stars. The PN, NGC 6302 is estimated to have an envelope with a total mass of 4.7 M$_{\sun}$ and a central star with a progenitor mass of 5.5 M$_{\sun}$ \citep{2011arXiv1107.4554W}.
 So the estimate made here for the total envelope mass of R~CrB is consistent with it being a fossil PN shell. 
 Any gas lost during the white-dwarf merger would have far less mass. 
 If the shell is an old PN shell then this would suggest that R~CrB is the product of a final flash event rather than a white-dwarf merger.
The estimated mass of the envelope is consistent with a mass-loss rate of 10$^{-5}$ M$_{\sun}$ for about 10$^{5}$ yr. 

 Five Galactic RCB stars, including R~CrB, itself, show significant Lithium in their atmospheres \citep{Rao:1996oq,Asplund:2000qy,Kipper:2006fk}. 
\citet{Renzini:1990wd} suggested that in a final flash the ingestion the burning of the H-rich envelope leads to Li-production through the Cameron-Fowler mechanism \citep{Cameron:1971lr}.
The abundance of Li in the atmosphere of the final-flash star, Sakurai's object, was actually observed to increase with time \citep{Asplund:1999bh}.
In a white-dwarf merger, the temperatures required for the production of $^{18}$O should destroy any Li present. 
So,  the simultaneous enrichment of Li and  $^{18}$O is not expected in the white-dwarf merger scenario. 
The abundance of $^{18}$O cannot be directly measured in R~CrB, but it  is overabundant in fluorine, which does imply a high $^{18}$O abundance \citep{Pandey:2008eu}.

About 10\% of single stars will undergo a final-flash event \citep{Iben:1996fj}.
About this percentage of RCB stars
(R~CrB, RY Sgr, V CrA, and UW Cen) show evidence of resolved fossil dust shells  \citep{walker94}. 
R~CrB is thought to be $\sim$0.8-0.9  M$_{\sun}$ from pulsation modeling \citep{Saio:2008qe}, and this mass agrees well with the predicted mass of the merger products of a CO- and a He-WD \citep{Han:1998kl}. On the other hand, final-flash stars, since they are single white dwarfs, should typically have masses of 0.55-0.6 M$_{\sun}$ \citep{Bergeron:2007tg}.

\section{Summary}
New images were obtained of R~CrB with {\it Gemini/GMOS}, {\it HST/WFPC2}, {\it Spitzer/MIPS}, and {\it Herschel/SPIRE} just before or during its recent very long and very deep decline. At visible wavelengths with {\it Gemini} and {\it HST}, the dust obscuring R~CrB acts as a natural coronagraph, allowing material close to the star to be imaged for the first time. In particular, several cometary knots have been detected within a few thousand AU of the star. 
The puffs of dust that cause declines in R~CrB and the cometary knots have similar inferred masses so they may be the same dust clouds seen in and out of the direct line of sight to the star. 
Also, the existence of a cluster of galaxies directly behind R~CrB was revealed for the first time. The IR images confirm the existence of a very large diffuse shell (radius $\sim$4 pc) around R~CrB that contains cool dust ($\sim$25 K) which is detected out to 500 \micron~and has a total mass of $\sim$2 M$_{\sun}$. This places a possible upper limit on the lifetime of R~CrB as an RCB of $\sim$10$^{5}$ yr. In addition, there is no significant mass of cold dust around R~CrB that was missed by {\it IRAS}. 

RCB stars are rare so they don't contribute significantly to the interstellar medium, but their continual dust formation makes them ideal laboratories for understanding dust formation in evolved stars. 
In particular, both the RCB stars and Type Ia supernovae may be the product of a white dwarf merger \citep{1978ApJ...225..212W,1984ApJ...277..355W,Clayton:2007ve,2010ApJ...716..122K}. So studies of the circumstellar environments of RCB stars may provide information on the material around Type Ia supernovae.
It is possible that there are two avenues for producing RCB stars and that R~CrB is a product of the final flash rather than the white-dwarf merger.
But the evidence is mixed. The IR shell and the presence of lithium point toward a possible final flash, while the relatively high mass of R~CrB and the presence of fluorine point to a WD merger.

\acknowledgments

We appreciate the granting of {\it Gemini} and {\it HST} Director's Discretionary time for this project.
This work is based in part on observations made with the {\it Spitzer Space Telescope}, which is operated by the Jet Propulsion Laboratory, California Institute of Technology under a contract with NASA. Support for this work was provided by NASA through an award (RSA No. 1287678) issued by JPL/Caltech. Support for Program number HST-GO-12000.01-A was provided by NASA through
a grant from the Space Telescope Science Institute, which is operated by the
Association of Universities for Research in Astronomy, Incorporated, under
NASA contract NAS5-26555. 
	A portion of these data was obtained at the {\it Gemini} Observatory, which is operated by the Association of Universities for Research in Astronomy (AURA) under a cooperative agreement with the NSF on behalf of the {\it Gemini} partnership. {\it Herschel} is an ESA space observatory with science instruments provided
by European-led Principal Investigator consortia and with important participation from NASA.
Funding for the SDSS and SDSS-II has been provided by the Alfred P. Sloan Foundation, the Participating Institutions, the National Science Foundation, the U.S. Department of Energy, the National Aeronautics and Space Administration, the Japanese Monbukagakusho, the Max Planck Society, and the Higher Education Funding Council for England. 
We acknowledge with thanks the variable star observations from the AAVSO International Database contributed by observers worldwide and used in this research.

\bibliography{/Users/gclayton/projects/latexstuff/everything2}

\begin{thebibliography}{82}
\expandafter\ifx\csname natexlab\endcsname\relax\def\natexlab#1{#1}\fi

\bibitem[{{Alcock et al.}(2001)}]{2001ApJ...554..298A}
{Alcock et al.} 2001, \apj, 554, 298

\bibitem[{{Asplund} {et~al.}(2000){Asplund}, {Gustafsson}, {Lambert}, \&
  {Rao}}]{Asplund:2000qy}
{Asplund}, M., {Gustafsson}, B., {Lambert}, D.~L., \& {Rao}, N.~K. 2000, \aap,
  353, 287

\bibitem[{{Asplund} {et~al.}(1999){Asplund}, {Lambert}, {Kipper}, {Pollacco},
  \& {Shetrone}}]{Asplund:1999bh}
{Asplund}, M., {Lambert}, D.~L., {Kipper}, T., {Pollacco}, D., \& {Shetrone},
  M.~D. 1999, \aap, 343, 507

\bibitem[{{Bergeron} {et~al.}(2007){Bergeron}, {Gianninas}, \&
  {Boudreault}}]{Bergeron:2007tg}
{Bergeron}, P., {Gianninas}, A., \& {Boudreault}, S. 2007, in A. S. P. Conf.
  Ser., Vol. 372, 15th European Workshop on White Dwarfs, ed. {R.~Napiwotzki \&
  M.~R.~Burleigh}, 29

\bibitem[{{Bright} {et~al.}(2011){Bright}, {Chesneau}, {Clayton}, {de Marco},
  {Le{\~a}o}, {Nordhaus}, \& {Gallagher}}]{2011MNRAS.414.1195B}
{Bright}, S.~N., {Chesneau}, O., {Clayton}, G.~C., {de Marco}, O., {Le{\~a}o},
  I.~C., {Nordhaus}, J., \& {Gallagher}, J.~S. 2011, \mnras, 414, 1195

\bibitem[{{Cameron} \& {Fowler}(1971)}]{Cameron:1971lr}
{Cameron}, A.~G.~W., \& {Fowler}, W.~A. 1971, \apj, 164, 111

\bibitem[{{Chevalier}(1986)}]{1986ApJ...308..225C}
{Chevalier}, R.~A. 1986, \apj, 308, 225

\bibitem[{{Clayton}(1996)}]{1996PASP..108..225C}
{Clayton}, G.~C. 1996, \pasp, 108, 225

\bibitem[{{Clayton} \& {Ayres}(2001)}]{2001ApJ...560..986C}
{Clayton}, G.~C., \& {Ayres}, T.~R. 2001, \apj, 560, 986

\bibitem[{{Clayton} {et~al.}(1997){Clayton}, {Bjorkman}, {Nordsieck},
  {Zellner}, \& {Schulte-Ladbeck}}]{1997ApJ...476..870C}
{Clayton}, G.~C., {Bjorkman}, K.~S., {Nordsieck}, K.~H., {Zellner}, N.~E.~B.,
  \& {Schulte-Ladbeck}, R.~E. 1997, \apj, 476, 870

\bibitem[{{Clayton} {et~al.}(2003){Clayton}, {Geballe}, \&
  {Bianchi}}]{2003ApJ...595..412C}
{Clayton}, G.~C., {Geballe}, T.~R., \& {Bianchi}, L. 2003, \apj, 595, 412

\bibitem[{{Clayton} {et~al.}(2007){Clayton}, {Geballe}, {Herwig}, {Fryer}, \&
  {Asplund}}]{Clayton:2007ve}
{Clayton}, G.~C., {Geballe}, T.~R., {Herwig}, F., {Fryer}, C., \& {Asplund}, M.
  2007, \apj, 662, 1220

\bibitem[{{Clayton} {et~al.}(2005){Clayton}, {Herwig}, {Geballe}, {Asplund},
  {Tenenbaum}, {Engelbracht}, \& {Gordon}}]{2005ApJ...623L.141C}
{Clayton}, G.~C., {Herwig}, F., {Geballe}, T.~R., {Asplund}, M., {Tenenbaum},
  E.~D., {Engelbracht}, C.~W., \& {Gordon}, K.~D. 2005, \apjl, 623, L141

\bibitem[{{Clayton} {et~al.}(1999){Clayton}, {Kerber}, {Gordon}, {Lawson},
  {Wolff}, {Pollacco}, \& {Furlan}}]{1999ApJ...517L.143C}
{Clayton}, G.~C., {Kerber}, F., {Gordon}, K.~D., {Lawson}, W.~A., {Wolff},
  M.~J., {Pollacco}, D.~L., \& {Furlan}, E. 1999, \apjl, 517, L143

\bibitem[{{Clayton} {et~al.}(1992){Clayton}, {Whitney}, {Stanford}, \&
  {Drilling}}]{1992ApJ...397..652C}
{Clayton}, G.~C., {Whitney}, B.~A., {Stanford}, S.~A., \& {Drilling}, J.~S.
  1992, \apj, 397, 652

\bibitem[{{Cottrell} {et~al.}(1990){Cottrell}, {Lawson}, \&
  {Buchhorn}}]{1990MNRAS.244..149C}
{Cottrell}, P.~L., {Lawson}, W.~A., \& {Buchhorn}, M. 1990, \mnras, 244, 149

\bibitem[{{Cox}(2000)}]{2000asqu.book.....C}
{Cox}, A.~N. 2000, {Allen's Astrophysical Quantities}

\bibitem[{{Crause} {et~al.}(2007){Crause}, {Lawson}, \&
  {Henden}}]{2007MNRAS.375..301C}
{Crause}, L.~A., {Lawson}, W.~A., \& {Henden}, A.~A. 2007, \mnras, 375, 301

\bibitem[{{de Laverny} \& {M{\'e}karnia}(2004)}]{de-Laverny:2004lr}
{de Laverny}, P., \& {M{\'e}karnia}, D. 2004, \aap, 428, L13

\bibitem[{{Diolaiti} {et~al.}(2000){Diolaiti}, {Bendinelli}, {Bonaccini},
  {Close}, {Currie}, \& {Parmeggiani}}]{2000A&AS..147..335D}
{Diolaiti}, E., {Bendinelli}, O., {Bonaccini}, D., {Close}, L., {Currie}, D.,
  \& {Parmeggiani}, G. 2000, \aaps, 147, 335

\bibitem[{{Dowell} {et~al.}(2010){Dowell}, {Pohlen}, {Pearson}, {Griffin},
  {Lim}, {Bendo}, {Benielli}, {Bock}, {Chanial}, {Clements}, {Conversi},
  {Ferlet}, {Fulton}, {Gastaud}, {Glenn}, {Grundy}, {Guest}, {King}, {Leeks},
  {Levenson}, {Lu}, {Morris}, {Nguyen}, {O'Halloran}, {Oliver}, {Panuzzo},
  {Papageorgiou}, {Polehampton}, {Rigopoulou}, {Roussel}, {Schneider},
  {Schulz}, {Schwartz}, {Shupe}, {Sibthorpe}, {Sidher}, {Smith}, {Swinyard},
  {Trichas}, {Valtchanov}, {Woodcraft}, {Xu}, \& {Zhang}}]{2010SPIE.7731E.101D}
{Dowell}, C.~D., {et~al.} 2010, in Society of Photo-Optical Instrumentation
  Engineers (SPIE) Conference Series, Vol. 7731, Society of Photo-Optical
  Instrumentation Engineers (SPIE) Conference Series

\bibitem[{{Feast}(1986)}]{1986ASSL..128..151F}
{Feast}, M.~W. 1986, IAU Colloq. 87, 128, p. 151

\bibitem[{{Feast} {et~al.}(1997){Feast}, {Carter}, {Roberts}, {Marang}, \&
  {Catchpole}}]{Feast:1997lr}
{Feast}, M.~W., {Carter}, B.~S., {Roberts}, G., {Marang}, F., \& {Catchpole},
  R.~M. 1997, \mnras, 285, 317

\bibitem[{{Fernie} \& {Lawson}(1993)}]{1993MNRAS.265..899F}
{Fernie}, J.~D., \& {Lawson}, W.~A. 1993, \mnras, 265, 899

\bibitem[{{Garcia-Hernandez} {et~al.}(2011){Garcia-Hernandez}, {Kameswara Rao},
  \& {Lambert}}]{2011arXiv1107.1185G}
{Garcia-Hernandez}, D.~A., {Kameswara Rao}, N., \& {Lambert}, D.~L. 2011,
  ArXiv1107.1185

\bibitem[{{Garc{\'{\i}}a-Hern{\'a}ndez}
  {et~al.}(2010){Garc{\'{\i}}a-Hern{\'a}ndez}, {Lambert}, {Kameswara Rao},
  {Hinkle}, \& {Eriksson}}]{2010ApJ...714..144G}
{Garc{\'{\i}}a-Hern{\'a}ndez}, D.~A., {Lambert}, D.~L., {Kameswara Rao}, N.,
  {Hinkle}, K.~H., \& {Eriksson}, K. 2010, \apj, 714, 144

\bibitem[{{Garc{\'{\i}}a-Segura} {et~al.}(2006){Garc{\'{\i}}a-Segura},
  {L{\'o}pez}, {Steffen}, {Meaburn}, \& {Manchado}}]{2006ApJ...646L..61G}
{Garc{\'{\i}}a-Segura}, G., {L{\'o}pez}, J.~A., {Steffen}, W., {Meaburn}, J.,
  \& {Manchado}, A. 2006, \apjl, 646, L61

\bibitem[{{Gillett} {et~al.}(1986){Gillett}, {Backman}, {Beichman}, \&
  {Neugebauer}}]{Gillett:1986cr}
{Gillett}, F.~C., {Backman}, D.~E., {Beichman}, C., \& {Neugebauer}, G. 1986,
  \apj, 310, 842

\bibitem[{{Gordon} {et~al.}(2005){Gordon}, {Rieke}, {Engelbracht}, {Muzerolle},
  {Stansberry}, {Misselt}, {Morrison}, {Cadien}, {Young}, {Dole}, {Kelly},
  {Alonso-Herrero}, {Egami}, {Su}, {Papovich}, {Smith}, {Hines}, {Rieke},
  {Blaylock}, {P{\'e}rez-Gonz{\'a}lez}, {Le Floc'h}, {Hinz}, {Latter},
  {Hesselroth}, {Frayer}, {Noriega-Crespo}, {Masci}, {Padgett}, {Smylie}, \&
  {Haegel}}]{2005PASP..117..503G}
{Gordon}, K.~D., {et~al.} 2005, \pasp, 117, 503

\bibitem[{{Griffin} {et~al.}(2008){Griffin}, {Dowell}, {Lim}, {Bendo}, {Bock},
  {Cara}, {Castro-Rodriguez}, {Chanial}, {Clements}, {Gastaud}, {Guest},
  {Glenn}, {Hristov}, {King}, {Laurent}, {Lu}, {Mainetti}, {Morris}, {Nguyen},
  {Panuzzo}, {Pearson}, {Pinsard}, {Pohlen}, {Polehampton}, {Rizzo}, {Schulz},
  {Schwartz}, {Sibthorpe}, {Swinyard}, {Xu}, \& {Zhang}}]{2008SPIE.7010E..80G}
{Griffin}, M., {et~al.} 2008, in Society of Photo-Optical Instrumentation
  Engineers (SPIE) Conference Series, Vol. 7010, Society of Photo-Optical
  Instrumentation Engineers (SPIE) Conference Series

\bibitem[{{Griffin} {et~al.}(2010){Griffin}, {Abergel}, {Abreu}, {Ade},
  {Andr{\'e}}, {Augueres}, {Babbedge}, {Bae}, {Baillie}, {Baluteau}, {Barlow},
  {Bendo}, {Benielli}, {Bock}, {Bonhomme}, {Brisbin}, {Brockley-Blatt},
  {Caldwell}, {Cara}, {Castro-Rodriguez}, {Cerulli}, {Chanial}, {Chen},
  {Clark}, {Clements}, {Clerc}, {Coker}, {Communal}, {Conversi}, {Cox},
  {Crumb}, {Cunningham}, {Daly}, {Davis}, {de Antoni}, {Delderfield}, {Devin},
  {di Giorgio}, {Didschuns}, {Dohlen}, {Donati}, {Dowell}, {Dowell}, {Duband},
  {Dumaye}, {Emery}, {Ferlet}, {Ferrand}, {Fontignie}, {Fox}, {Franceschini},
  {Frerking}, {Fulton}, {Garcia}, {Gastaud}, {Gear}, {Glenn}, {Goizel},
  {Griffin}, {Grundy}, {Guest}, {Guillemet}, {Hargrave}, {Harwit}, {Hastings},
  {Hatziminaoglou}, {Herman}, {Hinde}, {Hristov}, {Huang}, {Imhof}, {Isaak},
  {Israelsson}, {Ivison}, {Jennings}, {Kiernan}, {King}, {Lange}, {Latter},
  {Laurent}, {Laurent}, {Leeks}, {Lellouch}, {Levenson}, {Li}, {Li},
  {Lilienthal}, {Lim}, {Liu}, {Lu}, {Madden}, {Mainetti}, {Marliani}, {McKay},
  {Mercier}, {Molinari}, {Morris}, {Moseley}, {Mulder}, {Mur}, {Naylor},
  {Nguyen}, {O'Halloran}, {Oliver}, {Olofsson}, {Olofsson}, {Orfei}, {Page},
  {Pain}, {Panuzzo}, {Papageorgiou}, {Parks}, {Parr-Burman}, {Pearce},
  {Pearson}, {P{\'e}rez-Fournon}, {Pinsard}, {Pisano}, {Podosek}, {Pohlen},
  {Polehampton}, {Pouliquen}, {Rigopoulou}, {Rizzo}, {Roseboom}, {Roussel},
  {Rowan-Robinson}, {Rownd}, {Saraceno}, {Sauvage}, {Savage}, {Savini},
  {Sawyer}, {Scharmberg}, {Schmitt}, {Schneider}, {Schulz}, {Schwartz},
  {Shafer}, {Shupe}, {Sibthorpe}, {Sidher}, {Smith}, {Smith}, {Smith},
  {Spencer}, {Stobie}, {Sudiwala}, {Sukhatme}, {Surace}, {Stevens}, {Swinyard},
  {Trichas}, {Tourette}, {Triou}, {Tseng}, {Tucker}, {Turner}, {Vaccari},
  {Valtchanov}, {Vigroux}, {Virique}, {Voellmer}, {Walker}, {Ward}, {Waskett},
  {Weilert}, {Wesson}, {White}, {Whitehouse}, {Wilson}, {Winter}, {Woodcraft},
  {Wright}, {Xu}, {Zavagno}, {Zemcov}, {Zhang}, \&
  {Zonca}}]{2010A&A...518L...3G}
{Griffin}, M.~J., {et~al.} 2010, \aap, 518, L3

\bibitem[{{Groenewegen} {et~al.}(2011){Groenewegen}, {Waelkens}, {Barlow},
  {Kerschbaum}, {Garcia-Lario}, {Cernicharo}, {Blommaert}, {Bouwman}, {Cohen},
  {Cox}, {Decin}, {Exter}, {Gear}, {Gomez}, {Hargrave}, {Henning},
  {Hutsem{\'e}kers}, {Ivison}, {Jorissen}, {Krause}, {Ladjal}, {Leeks}, {Lim},
  {Matsuura}, {Naz{\'e}}, {Olofsson}, {Ottensamer}, {Polehampton}, {Posch},
  {Rauw}, {Royer}, {Sibthorpe}, {Swinyard}, {Ueta}, {Vamvatira-Nakou},
  {Vandenbussche}, {van de Steene}, {van Eck}, {van Hoof}, {van Winckel},
  {Verdugo}, \& {Wesson}}]{2011A&A...526A.162G}
{Groenewegen}, M.~A.~T., {et~al.} 2011, \aap, 526, 162

\bibitem[{{Han}(1998)}]{Han:1998kl}
{Han}, Z. 1998, \mnras, 296, 1019

\bibitem[{{Hecht} {et~al.}(1998){Hecht}, {Clayton}, {Drilling}, \&
  {Jeffery}}]{1998ApJ...501..813H}
{Hecht}, J.~H., {Clayton}, G.~C., {Drilling}, J.~S., \& {Jeffery}, C.~S. 1998,
  \apj, 501, 813

\bibitem[{{Hecht} {et~al.}(1984){Hecht}, {Holm}, {Donn}, \&
  {Wu}}]{1984ApJ...280..228H}
{Hecht}, J.~H., {Holm}, A.~V., {Donn}, B., \& {Wu}, C.-C. 1984, \apj, 280, 228

\bibitem[{{Henyey} \& {Greenstein}(1941)}]{1941ApJ....93...70H}
{Henyey}, L.~G., \& {Greenstein}, J.~L. 1941, \apj, 93, 70

\bibitem[{{Iben} {et~al.}(1996){Iben}, {Tutukov}, \& {Yungelson}}]{Iben:1996fj}
{Iben}, Jr., I., {Tutukov}, A.~V., \& {Yungelson}, L.~R. 1996, \apj, 456, 750

\bibitem[{{Ishihara} {et~al.}(2010){Ishihara}, {Onaka}, {Kataza}, {Salama},
  {Alfageme}, {Cassatella}, {Cox}, {Garc{\'{\i}}a-Lario}, {Stephenson},
  {Cohen}, {Fujishiro}, {Fujiwara}, {Hasegawa}, {Ita}, {Kim}, {Matsuhara},
  {Murakami}, {M{\"u}ller}, {Nakagawa}, {Ohyama}, {Oyabu}, {Pyo}, {Sakon},
  {Shibai}, {Takita}, {Tanab{\'e}}, {Uemizu}, {Ueno}, {Usui}, {Wada},
  {Watarai}, {Yamamura}, \& {Yamauchi}}]{2010A&A...514A...1I}
{Ishihara}, D., {et~al.} 2010, \aap, 514, A1

\bibitem[{{Jeffers et al.}(2011)}]{jeffers2011}
{Jeffers et al.}, J.~V. 2011, \aap, Submitted

\bibitem[{{Kilic} {et~al.}(2010){Kilic}, {Brown}, {Allende Prieto}, {Kenyon},
  \& {Panei}}]{2010ApJ...716..122K}
{Kilic}, M., {Brown}, W.~R., {Allende Prieto}, C., {Kenyon}, S.~J., \& {Panei},
  J.~A. 2010, \apj, 716, 122

\bibitem[{{Kipper} \& {Klochkova}(2006)}]{Kipper:2006fk}
{Kipper}, T., \& {Klochkova}, V.~G. 2006, Baltic Astronomy, 15, 531

\bibitem[{{Landolt} \& {Clem}(2010)}]{2010PASP..122..541L}
{Landolt}, A.~U., \& {Clem}, J.~L. 2010, \pasp, 122, 541

\bibitem[{{Lawson} {et~al.}(1990){Lawson}, {Cottrell}, {Kilmartin}, \&
  {Gilmore}}]{Lawson:1990fk}
{Lawson}, W.~A., {Cottrell}, P.~L., {Kilmartin}, P.~M., \& {Gilmore}, A.~C.
  1990, \mnras, 247, 91

\bibitem[{{Le{\~a}o} {et~al.}(2007){Le{\~a}o}, {de Laverny}, {Chesneau},
  {M{\'e}karnia}, \& {de Medeiros}}]{2007A&A...466L...1L}
{Le{\~a}o}, I.~C., {de Laverny}, P., {Chesneau}, O., {M{\'e}karnia}, D., \& {de
  Medeiros}, J.~R. 2007, \aap, 466, L1

\bibitem[{{Li} \& {Draine}(2001)}]{2001ApJ...554..778L}
{Li}, A., \& {Draine}, B.~T. 2001, \apj, 554, 778

\bibitem[{{Loreta}(1935)}]{1935AN....254..151L}
{Loreta}, E. 1935, Astronomische Nachrichten, 254, 151

\bibitem[{{Mathis} {et~al.}(1977){Mathis}, {Rumpl}, \&
  {Nordsieck}}]{1977ApJ...217..425M}
{Mathis}, J.~S., {Rumpl}, W., \& {Nordsieck}, K.~H. 1977, \apj, 217, 425

\bibitem[{{Matsuura} {et~al.}(2009){Matsuura}, {Speck}, {McHunu}, {Tanaka},
  {Wright}, {Smith}, {Zijlstra}, {Viti}, \& {Wesson}}]{2009ApJ...700.1067M}
{Matsuura}, M., {et~al.} 2009, \apj, 700, 1067

\bibitem[{{Miville-Desch{\^e}nes} \& {Lagache}(2005)}]{2005ApJS..157..302M}
{Miville-Desch{\^e}nes}, M., \& {Lagache}, G. 2005, \apjs, 157, 302

\bibitem[{{Murakami} {et~al.}(2007){Murakami}, {Baba}, {Barthel}, {Clements},
  {Cohen}, {Doi}, {Enya}, {Figueredo}, {Fujishiro}, {Fujiwara}, {Fujiwara},
  {Garcia-Lario}, {Goto}, {Hasegawa}, {Hibi}, {Hirao}, {Hiromoto}, {Hong},
  {Imai}, {Ishigaki}, {Ishiguro}, {Ishihara}, {Ita}, {Jeong}, {Jeong},
  {Kaneda}, {Kataza}, {Kawada}, {Kawai}, {Kawamura}, {Kessler}, {Kester},
  {Kii}, {Kim}, {Kim}, {Kobayashi}, {Koo}, {Kwon}, {Lee}, {Lorente}, {Makiuti},
  {Matsuhara}, {Matsumoto}, {Matsuo}, {Matsuura}, {M{\"u}ller}, {Murakami},
  {Nagata}, {Nakagawa}, {Naoi}, {Narita}, {Noda}, {Oh}, {Ohnishi}, {Ohyama},
  {Okada}, {Okuda}, {Oliver}, {Onaka}, {Ootsubo}, {Oyabu}, {Pak}, {Park},
  {Pearson}, {Rowan-Robinson}, {Saito}, {Sakon}, {Salama}, {Sato}, {Savage},
  {Serjeant}, {Shibai}, {Shirahata}, {Sohn}, {Suzuki}, {Takagi}, {Takahashi},
  {Tanab{\'e}}, {Takeuchi}, {Takita}, {Thomson}, {Uemizu}, {Ueno}, {Usui},
  {Verdugo}, {Wada}, {Wang}, {Watabe}, {Watarai}, {White}, {Yamamura},
  {Yamauchi}, \& {Yasuda}}]{2007PASJ...59S.369M}
{Murakami}, H., {et~al.} 2007, \pasj, 59, 369

\bibitem[{{O'Dell} \& {Handron}(1996)}]{1996AJ....111.1630O}
{O'Dell}, C.~R., \& {Handron}, K.~D. 1996, \aj, 111, 1630

\bibitem[{{Ohnaka} {et~al.}(2003){Ohnaka}, {Beckmann}, {Berger}, {Brewer},
  {Hofmann}, {Lacasse}, {Malanushenko}, {Millan-Gabet}, {Monnier}, {Pedretti},
  {Schertl}, {Schloerb}, {Shenavrin}, {Traub}, {Weigelt}, \&
  {Yudin}}]{2003A&A...408..553O}
{Ohnaka}, K., {et~al.} 2003, \aap, 408, 553

\bibitem[{{O'Keefe}(1939)}]{1939ApJ....90..294O}
{O'Keefe}, J.~A. 1939, \apj, 90, 294

\bibitem[{{Pandey} {et~al.}(2008){Pandey}, {Lambert}, \& {Rao}}]{Pandey:2008eu}
{Pandey}, G., {Lambert}, D.~L., \& {Rao}, N.~K. 2008, \apj, 674, 1068

\bibitem[{{Pigott} \& {Englefield}(1797)}]{1797RSPT...87..133P}
{Pigott}, E., \& {Englefield}, H.~C. 1797, Royal Society of London
  Philosophical Transactions Series I, 87, 133

\bibitem[{{Pilbratt} {et~al.}(2010){Pilbratt}, {Riedinger}, {Passvogel},
  {Crone}, {Doyle}, {Gageur}, {Heras}, {Jewell}, {Metcalfe}, {Ott}, \&
  {Schmidt}}]{2010A&A...518L...1P}
{Pilbratt}, G.~L., {et~al.} 2010, \aap, 518, L1

\bibitem[{{Pittard} {et~al.}(2005){Pittard}, {Dyson}, {Falle}, \&
  {Hartquist}}]{2005MNRAS.361.1077P}
{Pittard}, J.~M., {Dyson}, J.~E., {Falle}, S.~A.~E.~G., \& {Hartquist}, T.~W.
  2005, \mnras, 361, 1077

\bibitem[{{Pollacco} {et~al.}(1991){Pollacco}, {Hill}, {Houziaux}, \&
  {Manfroid}}]{1991MNRAS.248P...1P}
{Pollacco}, D.~L., {Hill}, P.~W., {Houziaux}, L., \& {Manfroid}, J. 1991,
  \mnras, 248, 1P

\bibitem[{{Rao} \& {Lambert}(1996)}]{Rao:1996oq}
{Rao}, N.~K., \& {Lambert}, D.~L. 1996, in Astronomical Society of the Pacific
  Conference Series, Vol.~96, Hydrogen Deficient Stars, ed. {C.~S.~Jeffery \&
  U.~Heber}, 43

\bibitem[{{Rao} \& {Lambert}(2010)}]{Kameswara-Rao:2010lr}
{Rao}, N.~K., \& {Lambert}, D.~L. 2010, in Recent Advances in Spectroscopy
  Theoretical, Astrophysical and Experimental Perspectives, ed.
  {R.~K.~Chaudhuri, M.~V.~Mekkaden, A.~V.~Raveendran, \& A.~Satya Narayanan },
  177

\bibitem[{{Rao} \& {Nandy}(1986)}]{1986MNRAS.222..357K}
{Rao}, N.~K., \& {Nandy}, K. 1986, \mnras, 222, 357

\bibitem[{{Renzini}(1990)}]{Renzini:1990wd}
{Renzini}, A. 1990, in Astronomical Society of the Pacific Conference Series,
  Vol.~11, Confrontation Between Stellar Pulsation and Evolution, ed.
  {C.~Cacciari \& G.~Clementini}, 549

\bibitem[{{Robitaille} {et~al.}(2006){Robitaille}, {Whitney}, {Indebetouw},
  {Wood}, \& {Denzmore}}]{2006ApJS..167..256R}
{Robitaille}, T.~P., {Whitney}, B.~A., {Indebetouw}, R., {Wood}, K., \&
  {Denzmore}, P. 2006, \apjs, 167, 256

\bibitem[{{Saio}(2008)}]{Saio:2008qe}
{Saio}, H. 2008, in Astronomical Society of the Pacific Conference Series, Vol.
  391, Hydrogen-Deficient Stars, ed. {A.~Werner \& T.~Rauch}, 69

\bibitem[{{Schlegel} {et~al.}(1998){Schlegel}, {Finkbeiner}, \&
  {Davis}}]{1998ApJ...500..525S}
{Schlegel}, D.~J., {Finkbeiner}, D.~P., \& {Davis}, M. 1998, \apj, 500, 525

\bibitem[{{Stanford} {et~al.}(1988){Stanford}, {Clayton}, {Meade}, {Nordsieck},
  {Whitney}, {Murison}, {Nook}, \& {Anderson}}]{1988ApJ...325L...9S}
{Stanford}, S.~A., {Clayton}, G.~C., {Meade}, M.~R., {Nordsieck}, K.~H.,
  {Whitney}, B.~A., {Murison}, M.~A., {Nook}, M.~A., \& {Anderson}, C.~M. 1988,
  \apjl, 325, L9

\bibitem[{{Swinyard} {et~al.}(2010){Swinyard}, {Ade}, {Baluteau}, {Aussel},
  {Barlow}, {Bendo}, {Benielli}, {Bock}, {Brisbin}, {Conley}, {Conversi},
  {Dowell}, {Dowell}, {Ferlet}, {Fulton}, {Glenn}, {Glauser}, {Griffin},
  {Griffin}, {Guest}, {Imhof}, {Isaak}, {Jones}, {King}, {Leeks}, {Levenson},
  {Lim}, {Lu}, {Makiwa}, {Naylor}, {Nguyen}, {Oliver}, {Panuzzo},
  {Papageorgiou}, {Pearson}, {Pohlen}, {Polehampton}, {Pouliquen},
  {Rigopoulou}, {Ronayette}, {Roussel}, {Rykala}, {Savini}, {Schulz},
  {Schwartz}, {Shupe}, {Sibthorpe}, {Sidher}, {Smith}, {Spencer}, {Trichas},
  {Triou}, {Valtchanov}, {Wesson}, {Woodcraft}, {Xu}, {Zemcov}, \&
  {Zhang}}]{2010A&A...518L...4S}
{Swinyard}, B.~M., {et~al.} 2010, \aap, 518, L4

\bibitem[{{Tisserand} {et~al.}(2009){Tisserand}, {Wood}, {Marquette}, {Afonso},
  {Albert}, {Andersen}, {Ansari}, {Aubourg}, {Bareyre}, {Beaulieu}, {Charlot},
  {Coutures}, {Ferlet}, {Fouqu{\'e}}, {Glicenstein}, {Goldman}, {Gould},
  {Gros}, {de Kat}, {Lesquoy}, {Loup}, {Magneville}, {Maurice}, {Maury},
  {Milsztajn}, {Moniez}, {Palanque-Delabrouille}, {Perdereau}, {Rich},
  {Schwemling}, {Spiro}, \& {Vidal-Madjar}}]{Tisserand:2009fj}
{Tisserand}, P., {et~al.} 2009, \aap, 501, 985

\bibitem[{{Ueta} {et~al.}(2000){Ueta}, {Meixner}, \&
  {Bobrowsky}}]{2000ApJ...528..861U}
{Ueta}, T., {Meixner}, M., \& {Bobrowsky}, M. 2000, \apj, 528, 861

\bibitem[{{van Marle} {et~al.}(2011){van Marle}, {Meliani}, {Keppens}, \&
  {Decin}}]{2011ApJ...734L..26V}
{van Marle}, A.~J., {Meliani}, Z., {Keppens}, R., \& {Decin}, L. 2011, \apjl,
  734, L26

\bibitem[{{Walker}(1985)}]{Walker:1985rr}
{Walker}, H.~J. 1985, \aap, 152, 58

\bibitem[{{Walker}(1986)}]{1986ASSL..128..407W}
{Walker}, H.~J. 1986, in Astrophysics and Space Science Library, Vol. 128, IAU
  Colloq. 87: Hydrogen Deficient Stars and Related Objects, ed. K.~{Hunger},
  D.~{Schoenberner}, \& N.~{Kameswara Rao}, 407

\bibitem[{{Walker}(1994)}]{walker94}
---. 1994, CCP7 Newsletter, 21, 40

\bibitem[{{Webbink}(1984)}]{1984ApJ...277..355W}
{Webbink}, R.~F. 1984, \apj, 277, 355

\bibitem[{{Weingartner} \& {Draine}(2001)}]{2001ApJ...548..296W}
{Weingartner}, J.~C., \& {Draine}, B.~T. 2001, \apj, 548, 296

\bibitem[{{Wheeler}(1978)}]{1978ApJ...225..212W}
{Wheeler}, J.~C. 1978, \apj, 225, 212

\bibitem[{{Whitney} {et~al.}(1993){Whitney}, {Balm}, \&
  {Clayton}}]{1993ASPC...45..115W}
{Whitney}, B.~A., {Balm}, S.~P., \& {Clayton}, G.~C. 1993, in ASP Conf. Ser.:
  Luminous High-Latitude Stars, ed. D.~D. {Sasselov}, Vol.~45, 115

\bibitem[{{Whitney} {et~al.}(2003{\natexlab{a}}){Whitney}, {Wood}, {Bjorkman},
  \& {Cohen}}]{2003ApJ...598.1079W}
{Whitney}, B.~A., {Wood}, K., {Bjorkman}, J.~E., \& {Cohen}, M.
  2003{\natexlab{a}}, \apj, 598, 1079

\bibitem[{{Whitney} {et~al.}(2003{\natexlab{b}}){Whitney}, {Wood}, {Bjorkman},
  \& {Wolff}}]{2003ApJ...591.1049W}
{Whitney}, B.~A., {Wood}, K., {Bjorkman}, J.~E., \& {Wolff}, M.~J.
  2003{\natexlab{b}}, \apj, 591, 1049

\bibitem[{{Whittet}(2003)}]{2003dge..conf.....W}
{Whittet}, D.~C.~B. 2003, {Dust in the galactic environment}, 2nd edn.
  (Bristol: Institute of Physics (IOP) Publishing)

\bibitem[{{Woitke} {et~al.}(1996){Woitke}, {Goeres}, \&
  {Sedlmayr}}]{1996A&A...313..217W}
{Woitke}, P., {Goeres}, A., \& {Sedlmayr}, E. 1996, \aap, 313, 217

\bibitem[{{Wright} {et~al.}(2011){Wright}, {Barlow}, {Ercolano}, \&
  {Rauch}}]{2011arXiv1107.4554W}
{Wright}, N.~J., {Barlow}, M.~J., {Ercolano}, B., \& {Rauch}, T. 2011,
  arXiv1107.4554W

\end{thebibliography}

\clearpage

\clearpage

\begin{deluxetable}{lccc}
\tablewidth{0pt} 
\tablecaption{New Observations}
\tablenum{1}
\tablehead{\colhead{Date}&
\colhead{Telescope}&
           \colhead{Instr.}&
           \colhead{Filters}}
\startdata
2007 April 14/15&{\it Spitzer}&MIPS&24, 70, 160 \micron\\
2009 March 23 & {\it Gemini-S}&GMOS&g\arcmin, r\arcmin, i\arcmin, z\arcmin, CaT\\
2009 April 29&{\it HST}&WFPC2&F555W, F814W\\
2009 December 27&{\it Herschel}&SPIRE&250, 350, 500 \micron\\
\enddata
\end{deluxetable}

\begin{deluxetable}{lll}
\tabletypesize{\scriptsize}
\tablewidth{3in} 
\tablecaption{R~CrB Photometry}
\tablenum{2}
\tablehead{\colhead{Band$^a$}&
           \colhead{Flux (Jy)}& \colhead{$\sigma$ (Jy)}
           }
\startdata
U& 4.53e+00&3.90e-02\\
B& 1.16e+01&1.07e-01\\
V&  1.79e+01&1.66e-01\\
R$_c$&2.05e+01&1.90e-01\\
I$_c$&2.07e+01&1.91e-01\\
J  &1.77e+01&1.65e-01\\
H & 1.43e+01&1.00e-01\\
K  &1.43e+01&1.33e-01\\
L&2.56e+01&2.36e-01\\
AKARI/9&5.30e+01&2.44e+00\\
AKARI/18&2.15e+01&2.90e-02\\
IRAS/12&3.89e+01&1.55e+00\\
IRAS/25&1.71e+01&6.84e-01\\
IRAS/60&3.94e+00&3.15e-01\\
MIPS/70&2.03e+00&3.4E-02\\
AKARI/90&1.49e+00&1.14e-01\\
IRAS/100&2.00e+00&1.60e-01\\
MIPS160&2.97E-01&9.36E-03\\
SPIRE/250  &7.81E-02&1.17E-02\\
SPIRE/350&3.40E-02&5.10E-03\\
SPIRE/500&1.25E-02&4.34E-03
\enddata
\tablenotetext{a}{Maximum light UBVRI photometry from \citet{1990MNRAS.244..149C}. JHKL photometry from  
\citet{Feast:1997lr}.}
\end{deluxetable}

\begin{deluxetable}{l c c c c c c}
\tablecaption{Cometary Knot Models}
\tablenum{3}
\tablewidth{0pt}
\tablehead{
\colhead{Knot} & \colhead{$F_{\rm F555W}$} & \colhead{$F_{\rm F814W}$} &
\colhead{Grain} & \colhead{$\theta$} & \colhead{$r$} & \colhead{$M_{gr}$} \\
\colhead{} &
\multicolumn{2}{c}{$10^{-18}$ erg~cm$^{-2}$~s$^{-1}$~\AA$^{-1}$} &
\colhead{Model} & \colhead{\degree} & \colhead{AU} &
\colhead{M$_{\sun}$}
}
\startdata
1 &   723 &  261 & 0.1 $\mu$m  & 110 & 1500 & $3.7\times 10^{-9}$\\
 &       &      & MRN         & 35 & 2500  & $1.6\times 10^{-8}$\\
2 &   45  &   12 & 0.1 $\mu$m  & 70 & 4700  & $1.2\times 10^{-9}$\\
3 &   32  &   4  & 0.01 $\mu$m & varies & N/A   & (1-3)$\times 10^{-7}$\\
 &       &      & 0.001 $\mu$m& varies & N/A  & (1-5)$\times 10^{-4}$\\
 &       &   6  & 0.1 $\mu$m  & 50 & 3000  & $2.5\times 10^{-10}$\\
Halo & 15600 & 10600 & 1.0 $\mu$m  & N/A & $\lesssim$400 & $2-4\times 10^{-8}$
\enddata
\end{deluxetable}

\end{document}